\documentstyle[12pt,psfig]{article}
\begin{document}

\newcommand{\be}{\begin{equation}}
\newcommand{\ee}{\end{equation}}
\newcommand{\bea}{\begin{eqnarray}}
\newcommand{\eea}{\end{eqnarray}}
\def\d{{\rm d}}
\def\R{{\bf R}}
\def\P{{\rm P}}
\def\g{{\sqrt{-g}}}
\def\tr{\mathop{\rm Tr}\nolimits}
\def\mn{{\mu\nu}}
\def\p#1{{\partial}_{#1}}
\def\mapright#1{\smash{\mathop{\longrightarrow}\limits^{#1}}}
\def\mapdown#1{\big\downarrow \rlap{$\vcenter
  {\hbox{$\scriptstyle#1$}}$}}
\def\su#1{{\rm SU}(#1)}
\def\so#1{{\rm SO}(#1)}
\def\u#1{{\rm U}(#1)}
\def\o#1{{\rm O}(#1)}
\def\Z{{\bf Z}}
\def\M{{\cal M}}
\def\susy{{\sc SUSY}~}
\def\SUSY{{\sc SUSY}~}

~

\begin{flushright} hep-th/0011110 \\ EFI-2000-45 \end{flushright}

\vspace{0.2in}
\begin{center}
{\Large  {\bf TASI Lectures: Cosmology for String Theorists}}

\vspace{0.2in}
Sean M.  Carroll
\vskip 0.2cm
{{\it Enrico Fermi Institute and Department of Physics\\
University of Chicago \\
5640 S.~Ellis Avenue, Chicago, IL 60637, USA}
\\
\small email: {\tt carroll@theory.uchicago.edu}} \\
\small web: {\tt http://pancake.uchicago.edu/{\~{}}carroll/}
\end{center}

\vskip 1truecm

\begin{abstract}

These notes provide a brief introduction to modern cosmology, focusing
primarily on theoretical issues.  Some attention is paid to aspects of
potential interest to students of string theory, on both sides of the
two-way street of cosmological constraints on string theory and
stringy contributions to cosmology.  Slightly updated version of
lectures at the 1999 Theoretical Advanced Study Institute at the
University of Colorado, Boulder.

\end{abstract}

\medskip

\vfill

\newpage

\tableofcontents

\newpage

\renewcommand{\baselinestretch}{1}
\baselineskip 18pt

{\it Those who think of metaphysics as the most unconstrained
or speculative of disciplines are misinformed; compared with
cosmology, metaphysics is pedestrian and unimaginative.}
\begin{flushright} -- Stephen Toulmin\footnote{``Metaphysics''
is the traditional philosophical designation for the search
for an underlying theory of the structure of reality.  The
contemporary reader is welcome to substitute ``string theory''.}
\cite{toulmin} \end{flushright}

\section{Introduction}
\label{sec:introduction}

String theory and cosmology are two of the most ambitious intellectual
projects ever undertaken.  The former seeks to describe all of the
elements of nature and their interactions in a single coherent
framework, while the latter seeks to describe the origin, evolution,
and structure of the universe as a whole.  It goes without saying that
the ultimate success of each of these two programs will necessarily
involve an harmonious integration of the insights and requirements of
the other.

At this point, however, the connections between cosmology and
string theory are still rather tenuous.  Indeed, one searches in vain
for any appearance of ``cosmology'' in the index of a fairly comprehensive 
introductory textbook on string theory \cite{polchinski}, and
likewise for ``string theory'' in the index of a fairly comprehensive
introductory textbook on cosmology \cite{peebles}.

These absences cannot be attributed to a lack of knowledge or
imagination on the part of the authors.  Rather, they are a reflection
of a desire to stick largely to those aspects of these subjects about
which we can speak with some degree of confidence (although in
cosmology, at least, not everyone is so timid \cite{kt,lindebook}).  
In cosmology we
have a very successful framework for discussing the evolution of the
universe back to relatively early times and high temperatures, which
however does not reach all the way to the Planck era where stringy
effects are expected to become important.  In string theory,
meanwhile, we have learned a great deal about the behavior of the
theory in certain very special backgrounds, which however do not
include (in any obvious way) the conditions believed to obtain in the
early universe.

Fortunately, there is reason to believe that this situation may
change in the foreseeable future.  In cosmology, new data
coming in from a variety of sources hold the promise of shedding
new light on the inflationary era that is widely believed to have
occurred in the early universe, and which may have served as a bridge
from a quantum-gravity regime to a classical spacetime.  
And in string theory, the last few
years have witnessed a number of new proposals for formulating
the theory in settings which were previously out of reach, and
there are great hopes for continued progress in this direction.
Furthermore, there is a reasonable expectation of significant
improvement in our understanding of particle physics beyond the
standard model, from upcoming accelerator experiments as well as
attempts to directly detect cosmological dark matter.

It is therefore appropriate for cosmologists and string theorists to
keep a close watch on each other's work over the next few years, and
this philosophy has guided the preparation of these lectures.  I have
attempted to explain the basic framework of the standard cosmological
model in a mostly conventional way, but with an eye to those aspects
which would be most relevant to the application of string theory to
cosmology.  (Since these lectures were delivered, several reviews
have appeared which discuss aspects of string theory most relevant to 
cosmology \cite{lwc,Banks:1999ay,Veneziano:2000pz,Easson:2000mj}.)
My goals are purely pedagogical, which means for example
that I have made no real attempt to provide an accurate historical
account or a comprehensive list of references, instead focusing on a
selection of articles from which a deeper survey of the literature can
be begun.  Alternative perspectives can be found in a number of
other recent reviews of cosmology \cite{Lyth:1993eu,Kolb:1994ad,
Brandenberger:1996kf,Ellis:1998ct,turnertyson,Lazarides:1998pz}.

(Note:  These lectures were first written and delivered in summer
1999.  I have added occasional references to subsequent developments
where they seemed indispensible, but have made no effort at a
thorough updating.)

\section{The contemporary universe}
\label{sec:contemporary}

\subsection{Friedmann-Robertson-Walker cosmology}

The great simplifying fact of cosmology is that the universe
appears to be homogeneous (the same at every point) and
isotropic (the same in every direction) along a preferred set
of spatial hypersurfaces \cite{Peebles:1991ch,davis}.
Of course homogeneity and isotropy are only approximate, but
they become increasingly good approximations on larger length
scales, allowing us to describe spacetime on cosmological scales by 
the Robertson-Walker metric:
\be
  ds^2 = -\d t^2 + a^2(t)\left[{{\d r^2}\over{1-kr^2}}
  + r^2(\d\theta^2 +\sin^2\theta\,\d\phi^2)\right]\ ,\label{rwmetric}
\ee
where the scale factor $a(t)$ describes the relative size
of spacelike hypersurfaces at different times, and the curvature
parameter $k$ is $+1$ for positively curved spacelike
hypersurfaces, $0$ for flat hypersurfaces, and $-1$ for
negatively curved hypersurfaces.  These possibilities are more 
informally known as ``closed'', ``flat'', and ``open'' universes,
in reference to the spatial topology, but there are problems 
with such designations.  First, the flat and negatively-curved
spaces may in fact be compact manifolds obtained by global
identifications of their noncompact relatives 
\cite{ll,css,weeks,lsgsb}.  Second, there
is a confusion between the use of ``open''/``closed'' to refer
to spatial topology and the evolution of the universe; if
such universes are dominated by matter or radiation, the 
negatively curved ones will expand forever and the positively
curved ones will recollapse, but more general sources of
energy/momentum will not respect this relationship.

A photon traveling through an expanding universe will undergo
a redshift of its frequency proportional to the amount of
expansion; indeed we often use the redshift $z$ as a way of
specifying the scale factor at a given epoch:
\be
  1+z = {{\lambda_{\rm obs}}\over{\lambda_{\rm emitted}}}
  = {{a_0}\over{a_{\rm emitted}}}\ ,
\ee
where a subscript $0$ refers here and below to the value of a
quantity in the present universe.

Einstein's equations relate the dynamics of the scale
factor to the energy-momentum tensor.
For many cosmological applications we can assume that the
universe is dominated by a perfect fluid, in which case the
energy-momentum tensor is specified by an energy density $\rho$
and pressure $p$:
\be
  T_{00} = \rho\ ,\qquad T_{ij} = pg_{ij}\ ,
\ee
where indices $i$, $j$ run over spacelike values $\{1,2,3\}$.
The quantities $\rho$ and $p$ will be related by an equation
of state; many interesting fluids satisfy the simple equation
of state
\be
  p=w\rho\ ,
\ee
where $w$ is a constant independent of time.  The conservation
of energy equation $\nabla_\mu T^{\mu\nu}=0$ then implies
\be
  \rho \propto a^{-n}\ ,
\ee
with $n=3(1+w)$.  Especially popular equations of state include
the following:
\be
  \matrix{\rho \propto a^{-3} & \leftrightarrow & p=0 & \leftrightarrow & 
  {\rm matter,}\cr
  \rho \propto a^{-4} & \leftrightarrow & p={1\over 3}\rho & \leftrightarrow 
  &  {\rm radiation,}\cr
  \rho\propto a^0 & \leftrightarrow & p=-\rho & \leftrightarrow & 
  {\rm vacuum.}\cr}
\ee
``Matter'' (also called ``dust'') is used by cosmologists to refer to any 
set of non-relativistic, non-interacting particles; the pressure is
then negligible, and the energy density is dominated by the rest mass
of the particles, which redshifts away as the volume increases.
``Radiation'' includes any species of relativistic particles, for
which the individual particle energies will redshift as $1/a$ in
addition to the volume dilution factor.  (Coherent electromagnetic
fields will also obey this equation of state.)  The vacuum energy
density, equivalent to a cosmological constant $\Lambda$ via
$\rho_\Lambda = \Lambda /8\pi G$, is by definition
the energy remaining when all other forms of energy and momentum
have been cleared away.

Plugging the Robertson-Walker metric into Einstein's equations
yields the Friedmann equations,
\be
  \left({{\dot a}\over a}\right)^2={{8\pi G}\over 3}\rho
  -{{k}\over a^2} \label{friedmann}
\ee
and
\be
  {{\ddot a}\over a}=-{{4\pi G}\over 3}(\rho+3p) \ .
\ee
If the dependence of $\rho$ on the scale factor is known, equation
(\ref{friedmann}) is sufficient to solve for $a(t)$.

There is a host of terminology which is associated with the
cosmological parameters, and I will just introduce the basics
here.  The rate of expansion is characterized by the Hubble
parameter,
\be
  H ={{\dot a}\over a}\ .
\ee
The value of the Hubble parameter at the present epoch is the
Hubble constant, $H_0$. 
Another useful quantity is the density parameter in a species $i$,
\be
  \Omega_i =  {{8\pi G}\over {3H^2}}\rho_i
  = {{\rho_i}\over{\rho_{\rm crit}}}\ ,
\ee
where the critical density is defined by
\be
  \rho_{\rm crit} = {{3H^2}\over{8\pi G}}\ ,
\ee
corresponding to the energy density of a flat universe.  In
terms of the total density parameter
\be
  \Omega = \sum_i \Omega_i\ ,
\ee
the Friedmann equation  (\ref{friedmann}) can be written
\be
  \Omega-1={{k}\over {H^2 a^2}} \ .
\ee
The sign of $k$ is therefore determined by whether $\Omega$ is
greater than, equal to, or less than one.  We have
\[
  \matrix{\rho<\rho_{\rm crit} & \leftrightarrow & \Omega < 1 &
  \leftrightarrow & k=-1 & \leftrightarrow & {\rm open}\cr
  \rho=\rho_{\rm crit} & \leftrightarrow & \Omega = 1 &
  \leftrightarrow & k=0 & \leftrightarrow & {\rm flat}\cr
  \rho>\rho_{\rm crit} & \leftrightarrow & \Omega > 1 &
  \leftrightarrow & k=+1 & \leftrightarrow & {\rm closed}.\cr}
\]
\begin{figure}[t]
  \vskip-1.5cm
  \centerline{
  \psfig{figure=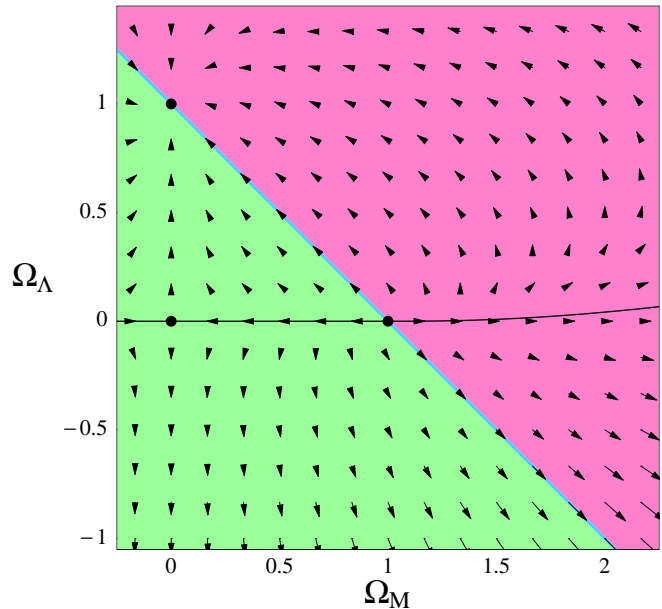,angle=0,height=12cm}}
  \vskip-1.5cm
  \caption{Dynamics of expanding universes dominated by 
  matter and vacuum energy.  The arrows indicate the direction
  of evolution.  Above and on the nearly-horizontal line are those
  universes which expand forever, while those below will
  eventually recollapse.}
  \label{ovecpale}
\end{figure}

\noindent
Note that $\Omega_i/\Omega_j = \rho_i/\rho_j = a^{-(n_i-n_j)}$,
so the relative amounts of energy in different components will
change as the universe evolves.  Figure \ref{ovecpale} 
shows how $\Omega_{\rm M}$
and $\Omega_{\Lambda}$ evolve in a universe dominated by matter and
a cosmological constant.  Note that the only attractive fixed 
point on the diagram is $(\Omega_{\rm M}=0, \Omega_\Lambda = 1)$.
In a sense this point represents the only natural stable solution 
for cosmology, and one of the outstanding problems is why we don't
find ourselves living there.

\subsection{Exact solutions}

Our actual universe consists of a complicated
stew of radiation, matter, and vacuum energy, as will be discussed
below.  It is nevertheless useful to consider exact solutions in
order to develop some intuition for cosmological dynamics.
The simplest solutions are those for flat universes, those with
$k=0$.  For flat universes it is often more convenient to use
Cartesian coordinates on spacelike hypersurfaces, so that the metric
takes the form 
\be
  ds^2 = -\d t^2 + a^2(t)[\d x^2 + \d y^2 + \d z^2] \ ,
\ee
rather than
the polar coordinates in (\ref{rwmetric}).  However, the solutions
for $a(t)$ are the same.  In a
flat universe dominated by a single energy
density source, the scaling of the source is directly related 
to the expansion history:
\be
  k=0,\  \rho \propto a^{-n} \ \rightarrow\ a \propto t^{2/n}\ .
\ee
(For $n=0$, we get exponential growth.)
Thus, a matter-dominated flat universe expands as
$a\propto t^{2/3}$, and a radiation-dominated flat universe as 
$a\propto t^{1/2}$.  When $k\neq 0$, the solutions for matter- and
radiation-dominated universes are slightly more complicated, but
may still be expressed in closed form 
\cite{HE,Carroll:1997ar}.  (Even when $k\neq 0$, the curvature
term $-k/a^2$ in the Friedmann equation will be subdominant to the
energy density for very small $a$, so it is sensible to model the
early universe using the $k=0$ metric.) 
When the energy density consists solely of matter and/or radiation
(or more generally when the energy density diminishes at least
as rapidly as $a^{-2}$), negatively curved universes
expand forever, while positively curved universes eventually
recollapse.  An interesting special case occurs for $\rho=0$, an
empty universe, which from (\ref{friedmann}) implies $k=-1$.  The
solution is then linear expansion, $a \propto t$; this is 
sometimes called the ``Milne universe''.  In fact the curvature
tensor vanishes in this spacetime, and it is simply an unconventional
coordinate system which covers a subset of Minkowski spacetime.

Let us now consider universes with a nonvanishing vacuum energy, with
all other energy set to zero.  Unlike ordinary energy, a cosmological
constant $\Lambda = 8\pi G\rho_\Lambda$ 
can be either positive or negative.  When $\Lambda >0$,
we can find solutions for any spatial curvature:
\be
  \matrix{k = -1 & \rightarrow & 
  a \propto \sinh\left[(\Lambda/3)^{1/2} \ t \right]  \cr
  k=0 & \rightarrow & 
  a \propto \exp\left[(\Lambda/3)^{1/2} \ t \right]  \cr
  k=+1 & \rightarrow & 
  a \propto \cosh\left[(\Lambda/3)^{1/2} \ t \right]  .\cr}
\ee
In fact, all of these solutions are the same spacetime, ``de~Sitter
space'', just expressed in different coordinates.  Any given 
spacetime can (locally) be foliated into spacelike hypersurfaces in 
infinitely many ways, although typically such hypersurfaces will be wildly
inhomogeneous; de~Sitter space has the property that it admits
Robertson-Walker foliations with any of the three spatial 
geometries (just as Minkowski space can be foliated either by surfaces
of constant negative curvature to obtain the Milne universe, or more
conventionally by flat hypersurfaces).
In general such foliations will not cover the entire spacetime:
the $k=+1$ coordinates cover all of de~Sitter (which has global
topology $\R\times S^3$), while the others
do not.  However, this doesn't imply that the $k=+1$ RW metric
is a ``better'' representation of de~Sitter.  For different
purposes, it might be useful to model a patch of some spacetime
by a patch of de~Sitter in certain coordinates.  For example, if
our universe went through an early phase in which it was dominated
by a large positive vacuum energy (as in the inflationary
scenario, discussed below),  but containing some trace test particles,
it would be natural to choose a coordinate system in which the
particles were comoving (traveling on worldlines orthogonal to
hypersurfaces of constant time), which might be the flat or
negatively-curved representations.
See Hawking and Ellis \cite{HE} for a discussion of
the connections between different coordinate systems.

When $\Lambda<0$, equation (\ref{friedmann}) implies that the
universe must have $k=-1$.  For this case the solution is
\be
  a \propto \sin\left[(-\Lambda/3)^{1/2} \ t \right] \ .
\ee
This universe is known as ``anti-de~Sitter space'', or ``AdS''
for short.  The RW 
coordinates describe an open universe which expands from a 
Big Bang, reaches a maximum value of the scale factor, and
recontracts to a Big Crunch (recall that for a nonzero $\Lambda$
the traditional relationship between spatial curvature and
temporal evolution does not hold).  Again, however,
these coordinates do not cover the entire spacetime (which has
global topology $\R^4$).  There are
a number of different coordinates that are useful on AdS, and
they have been much explored by string theorists in the context of
the celebrated correspondence between string theory on AdS in
$n$ dimensions and conformal field theory in $n-1$ dimensions;
see \cite{Aharony:2000ti} for a discussion.
One of the reasons why AdS plays a featured role in string theory
is that unbroken supersymmetry implies that the cosmological
constant is either negative or zero (see \cite{weinberg,carroll00}
and references therein).  Of course, in our low-energy world
supersymmetry is broken if it exists at all, and \susy breaking
generally contributes a positive vacuum energy, so one might 
think that it is not so surprising that we observe a positive
cosmological constant (see below).  The surprise is more
quantitative; the scale of \susy breaking is at least
$10^3$~GeV, while that of the vacuum energy is $10^{-12}$~GeV.

de~Sitter and anti-de~Sitter, along with Minkowski space, have the
largest possible number of isometries for a Lorentzian manifold
of the appropriate dimension; they are therefore known as
``maximally symmetric'' (and are the only such spacetimes).
In an $n$-dimensional maximally symmetric space, the Riemann 
tensor satisfies
\be
  R_{\mu\nu\rho\sigma} = {1\over n(n-1)}R(g_{\mu\rho}g_{\nu\sigma}
  - g_{\mu\sigma}g_{\nu\rho})\ ,
\ee
where $R$ is the Ricci scalar, which in this case is constant over
the entire manifold.
The well-known symmetries of Minkowski space
include the Lorentz group $\so{n-1,1}$ and the translations
$\R^4$, together known as the Poincar\'e group.  de~Sitter space
possesses an $\so{n,1}$ symmetry, while AdS has an $\so{n-1,2}$
symmetry.  All of these groups are of dimension $n(n+1)/2$.
There is a sense in which the maximally symmetric solutions can
be thought of as ``vacua'' of general relativity.  In the presence
of dynamical matter and energy (or gravitational waves), 
the solution will be non-vacuum, and possess less symmetry.

\subsection{Matter}

An inventory of the constituents comprising the actual universe
is hampered somewhat by the fact that they are not all equally
visible.  The first things we notice are
galaxies: collections of self-gravitating stars, 
gas, and dust.  The light from distant galaxies is (almost always)
redshifted, and the apparent recession velocity depends (almost 
exactly) linearly on distance:  $v = H_0 d$, where we 
interpret the slope as the Hubble parameter at the present
epoch.  (The ``almost''s are inserted because galaxies are not
perfectly comoving objects, but have proper motions that lead
to the conventional Doppler shifting; not to mention that at
sufficiently large distances the linear Hubble law will break down.)
Measuring extragalactic distances is notoriously tricky, but
most current measurements of the Hubble constant are consistent
with $H_0 = 60-80$~km/sec/Mpc, where 1~Mpc $= 10^6$~parsecs $=3\times
10^{24}$~cm \cite{freedman}.  In particle-physics units 
($\hbar = c = 1$), this is $H_0 \sim 10^{-33}$~eV.  It is
convenient to express the Hubble constant as $H_0 = 100h$~km/sec/Mpc,
where $0.6\leq h \leq 0.8$.  Note that, since 
$\rho_i = 3H_0^2\Omega_i/8\pi G$, measurements of $\rho_i$
will often be expressed as measurements of $\Omega_i h^2$.
The Hubble constant provides a rough measure of the scale of the
universe, since the age of a matter- or radiation-dominated 
universe is $t_0 \sim H_0^{-1}$.

We find perhaps $10^{11}$ stars in a typical galaxy.  The total
amount of luminous matter in all the galaxies we see adds up
to approximately $\Omega_{\rm lum}\sim 10^{-3}$.
In fact, most of the baryons are not in the form of stars,
but in ionized gas; our best estimates of the total baryon
density yield $\Omega_{\rm B}\sim 2\times 10^{-2}$ 
\cite{sal1,Fukugita:1997bi}.
But the dynamics of individual galaxies implies that there
is even more matter there, in ``halos'' \cite{sal2}.  The implied 
existence of ``dark matter'' is confirmed by applying the
virial theorem to clusters of galaxies, by looking at the
temperature profiles of clusters, by ``weighing'' clusters
using gravitational lensing, and by the large-scale motions
of galaxies between clusters.  The overall impression is
of a matter density corresponding to $\Omega_{\rm M} \sim 0.1-0.4$
\cite{bahcall,primack,freedman,Turner:1999kz,carroll00}.

There are innumerable fascinating facts
about the matter in the universe.
First and arguably foremost, it all seems to be matter and not antimatter
\cite{Cohen:1997ac}.  If, for 
example, half of the galaxies we observe were composed 
completely of antimatter, we would expect to see copious 
$\gamma$-ray emission from proton-antiproton annihilation in the
gas in between the galaxies.  Since it seems more natural to
imagine initial conditions in which matter and antimatter were
present in equal abundances, it appears necessary to invoke a 
dynamical mechanism to generate the observed asymmetry, as will
be discussed briefly in section (\ref{subsec:thermhist}).

The relative abundances of various elements are also of 
interest.  Heavy elements can be produced in stars, but it
is possible to deduce ``primordial'' abundances through careful
observation.  Most of the primordial baryons in the universe are to
be found in the form of hydrogen, with about $25\%$ helium-4 (by mass),
between $10^{-5}$ and $10^{-4}$ in deuterium, about $10^{-5}$
in helium-3, and $10^{-10}$ in lithium.  As discussed
below, these abundances provide a sensitive probe of 
early-universe cosmology \cite{schrammturner,sarkar96,mendozahogan}.

Besides baryons and dark matter, galaxies also possess
large-scale magnetic fields with root-mean-square amplitudes of 
order $10^{-6}$~Gauss \cite{zh,olinto}.
These fields may be the result of
dynamo amplification of small seed fields created early in
the history of the galaxies, or they may be relics of 
processes at work in the very early universe.

Finally, we have excellent evidence for the existence of
black holes in galaxies.  There are black holes of several
solar masses which are thought to be the end-products of the 
lives of massive stars, as well as supermassive black holes
($M\geq 10^6 M_\odot$) at the centers of galaxies
\cite{krolik,richstone}.  In
astrophysical situations the electric charges of black
holes will be negligible compared to their mass, since 
any significant charge will be quickly neutralized by
absorbing oppositely charged particles from the surrounding
plasma.  They can, however, have significant spin, and
observations have tentatively indicated spin parameters
$a\geq 0.95$ (where $a=1.0$ in an extremal Kerr black hole)
\cite{kerr}. 

\subsection{Cosmic Microwave Background}
\label{subsec:cmb}

Besides the matter (luminous and dark) found in the universe,
we also observe diffuse photon backgrounds \cite{ressellturner}.  
These come in all wavelengths, but most of the photons 
are to be found
in a nearly isotropic background with a thermal spectrum
at a temperature \cite{fixsen}
of $2.73~^\circ$K --- the cosmic microwave
background.  Careful observation has failed to find any
deviation from a perfect blackbody curve; indeed,
the CMB spectrum as measured by the COBE satellite is the
most precisely measured blackbody curve in all of physics.

Why does the spectrum have this form?  Typically, blackbody
radiation is emitted by systems in thermal equilibrium.
Currently, the photon background is essentially non-interacting,
and there is no accurate sense in which the universe is in
thermal equilibrium.  However, as the universe expands,
individual photon frequencies redshift with $\nu\propto 1/a$, and
a blackbody curve will be preserved, with temperature
$T\propto 1/a$.  Since the universe is expanding now, it
used to be smaller, and the temperature correspondingly
higher.  At sufficiently high temperatures the photons were
frequently interacting; specifically, at temperatures above
approximately 13~eV, hydrogen was ionized, and the photons
were coupled to charged particles.  The moment when the
temperature become low enough for hydrogen to be stable 
(at a redshift of order $10^3$) the universe became transparent.
This moment is known as ``recombination'' or ``decoupling'',
and the CMB we see today is to a good approximation a
snapshot of the universe at this epoch\footnote{Occasionally
a stickler will complain that ``recombination'' is a misnomer,
since the electrons are combining with protons for the first
time.  Such people should be dealt with by pointing out that
a typical electron will combine and dissociate with a proton
many times before finally settling down, so ``re-'' is a
perfectly appropriate prefix in describing the last of 
these combinations.}.

Today, there are 422 CMB photons per cubic centimeter,
which leads to a density parameter $\Omega_{\rm CMB}\sim 5\times
10^{-5}$.  If neutrinos are massless (or sufficiently light), a 
hypothetical neutrino background should contribute an energy
density comparable to that in photons.  We don't know of any
other significant source of energy density in radiation, so in
the contemporary universe the radiation energy density is
dominated by the matter energy density.  But of course they
depend on the scale factor in different ways, such that
$\Omega_{\rm M}/\Omega_{\rm R}\propto  a$.  Thus, matter-radiation
equality should have occurred at a redshift $z_{\rm EQ}\sim 10^4
\Omega_{\rm M}$.

The source of most current interest in the CMB is the small
but crucial temperature anisotropies from point to point in
the sky \cite{Hu:1997qs,Kamionkowski:1999qc,Barreiro:1999ct}.  
We typically decompose the temperature fluctuations
into spherical harmonics,
\be
  {{\Delta T}\over T} = \sum_{lm} a_{lm} Y_{lm}(\theta,\phi)\ ,
\ee
and express the amount of anisotropy at multipole moment $l$ via 
the power spectrum,
\be
  C_l = \langle |a_{lm}|^2 \rangle\ .
\ee
Higher multipoles correspond to smaller angular separations
on the sky, $\theta = 180^\circ/l$.

\begin{figure}[t]
  \centerline{
  \psfig{figure=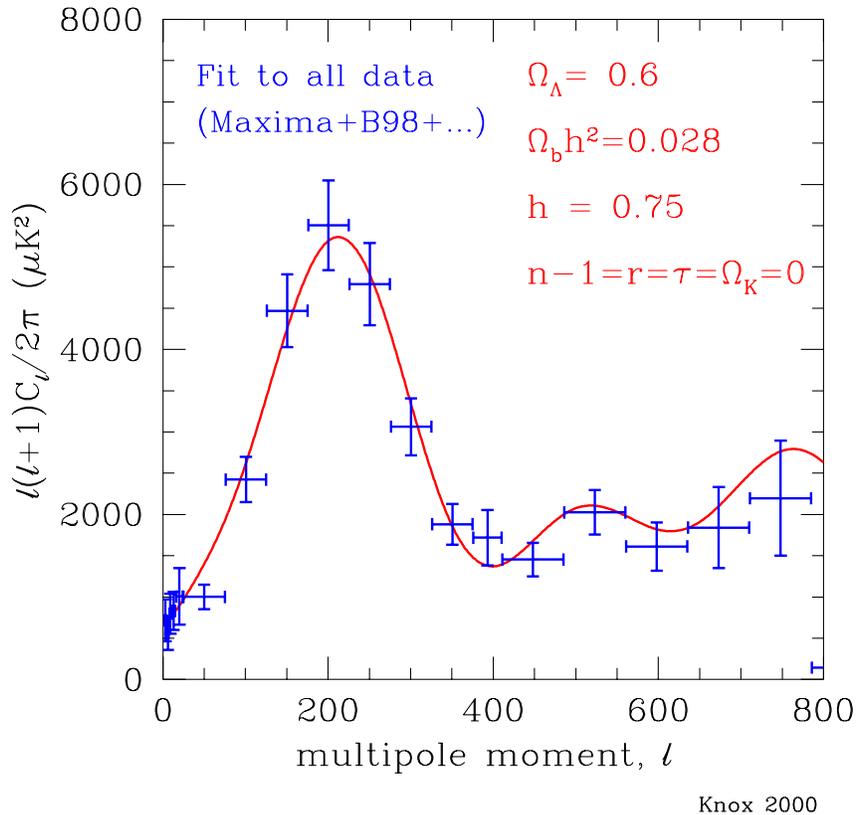,angle=0,height=12cm}}
  \caption{Amplitude of CMB temperature anisotropies, as a function
  of multipole moment $l$ (so that angular scale decreases from
  left to right).  The data points are averaged from all of the
  experiments performed as of Summer 2000.  The curve is a theoretical
  model with scale-free adiabatic scalar perturbations in a flat
  universe dominated by a cosmological constant.
  Courtesy of Lloyd Knox.} 
  \label{knoxcmb}
\end{figure}

Figure \ref{knoxcmb} shows a summary of data as of 
summer 2000, with various
experimental results consolidated into bins, along with a
theoretical model.  (See \cite{toco,boomna,boom,max}
for some recent observational work.)
The curve shown in the figure is based on currently a specific
understanding of the primordial inhomogeneities, in which
they are Gaussian fluctuations of approximately equal
magnitudes at all length scales (a ``Harrison-Zeldovich spectrum'') 
in a cold dark
matter component, which are ``adiabatic'' in the sense that 
fluctuations in the dark matter, photons, and baryons are all
correlated with each other.  A member of this family of models is
characterized by cosmological parameters such as the Hubble
constant, the $\Omega_i$'s, and the amplitude of the initial
fluctuations.  A happy feature of these models is the existence
of ``acoustic peaks'' in the CMB spectrum, whose characteristics
are closely tied to the cosmological parameters.  The first
peak (the one at lowest $l$) corresponds to the angular scale
subtended by the Hubble radius $H_{\rm CMB}^{-1}$ at 
recombination, which we can understand 
in simple physical terms \cite{Hu:1997qs}.

An overdense region of a given size $R$ will contract under the
influence of its own gravity, which occurs over a timescale
$\sim R$ (remember $c=1$).  For scales $R \gg H_{\rm CMB}^{-1}$,
overdense regions will not have had time to collapse in the
lifetime of the universe at last scattering.  For 
$R \leq H_{\rm CMB}^{-1}$, protons and electrons will have had
time to fall into the gravitational potential wells, raising
the temperature in the overdense regions (and lowering it in
the underdense ones).  There will be a restoring force due to
the increased photon pressure, leading to acoustic oscillations
which are damped by photon
diffusion.  The maximum amount of temperature anisotropy occurs
on the scale which has just had time to collapse but not
equilibrate, $R\sim H_{\rm CMB}^{-1}$, which appears to us
as a peak in the CMB anisotropy spectrum.

The angular scale at which we observe this peak is tied to the
geometry of the universe:  in a negatively (positively)
curved universe, photon paths diverge (converge), leading to 
a larger (smaller) apparent angular size as compared to a
flat universe \cite{Hu:1996qz,wcmb}.  Although the evolution of the
scale factor also influences the observed angular scale, for
reasonable values of the parameters this effect cancels out
and the location of the first peak will depend primarily on
the geometry.  In a flat universe, we have
\be
  l_{\rm peak} \sim 200\ ;
\ee
negative curvature moves the peak to higher $l$, and 
positive curvature to lower $l$.  It is clear from the figure
that this is indeed the observed location of the peak; this
result is the best evidence we have that we live in a flat
($k=0$, $\Omega=1$) Robertson-Walker universe.

More details about the spectrum (height of the peak, features
of the secondary peaks) will depend on other cosmological
quantities, such as the Hubble constant and the baryon
density.  Combined with constraints from other sources, data
which will be gathered in the near future from new satellite,
balloon and ground-based experiments should provide a wealth
of information that will help pin down the parameters
describing our universe.  You can calculate the theoretical
curves at home yourself with the program
CMBFAST \cite{cmbfast}.  The CMB can also be used to
constrain particle physics in various ways 
\cite{Kamionkowski:1999qc}.

\subsection{Evolution of the scale factor}
\label{subsec:scalefactor}

Saul Perlmutter's lectures at TASI-99 
discussed the recent observations of Type Ia
supernovae as standard candles, and the surprising result that
they seem to indicate an accelerating universe and therefore
a nonzero cosmological constant (or close relative thereof)
\cite{riess,perlmutter,Sahni:1999gb,Turner:1998hj}.  
Since wonderfully entertaining reviews have recently become
available \cite{carroll00}, I will not go into any detail 
here about this result and its consequences.  The important
point is that the supernova results have received confirmation
from a combination of dynamical measurements of $\Omega_{\rm M}$
and the CMB constraints on $\Omega_{\rm tot}$ discussed in
the previous section.  The favored universe is
one with $\Omega_{\rm M}\sim 0.3$ and $\Omega_\Lambda\sim 0.7$.

\begin{figure}[t]
  \centerline{
  \psfig{figure=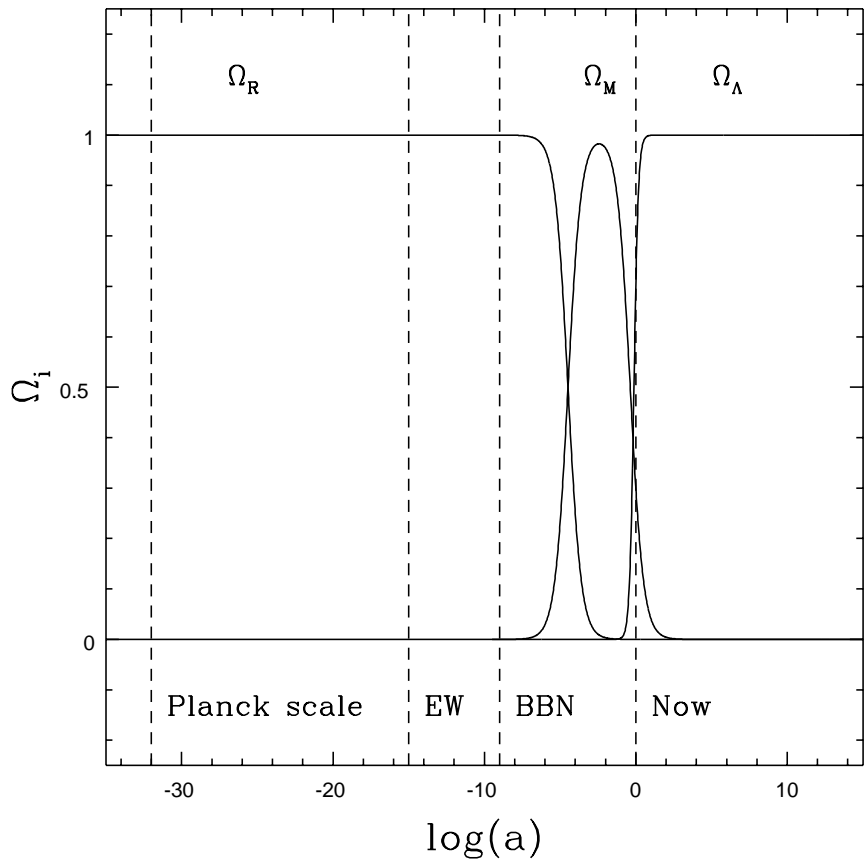,angle=0,height=12cm}}
  \caption{Evolution of the different density parameters
  in a universe with $\Omega_{0,{\rm M}}=0.3$, $\Omega_{0,\Lambda}
  =0.7$, and $\Omega_{0,{\rm R}}=5\times 10^{-5}$.}
  \label{oplotall}
\end{figure}

If true, this is a remarkable universe, especially considering
our early remark that the different $\Omega_i$'s evolve at
different rates.  Figure \ref{oplotall} shows this evolution for the
apparently-favored universe, as a function of log$(a)$.
The period in which $\Omega_{\rm M}$ is of the same order as
$\Omega_\Lambda$ is a very brief one, cosmically speaking.
It's clearly crucial that we work to better understand this
remarkable result, which will have important consequences for
both cosmology and fundamental microphysics if it is eventually
confirmed \cite{cpt,weinberg,carroll00}.

\section{The youthful universe}
\label{sec:youthful}

\subsection{Starting point}

In the previous lecture we discussed the universe as we see it,
as well as the dynamical equations which describe its evolution
according to general relativity.  One conclusion is that the
very early universe was much smaller and hotter than the universe
today, and the energy density was radiation-dominated.  We also
saw that the universe on large scales could be accurately
described by a perturbed Robertson-Walker metric.  On thermodynamic
grounds (backed up by evidence from CMB anisotropy)
it seems likely that these perturbations are growing
rather than shrinking with time, at least in the matter-dominated
era; it would require extreme fine-tuning of initial conditions
to arrange for diminishing matter perturbations
\cite{collinshawking}.  Thus, the early universe was smoother
as well.

Let us therefore trace the history of the universe as we 
reconstruct it given these conditions plus our current best
guesses at the relevant laws of physics.  We can start at a
temperature close to but not quite at the reduced\footnote{The
``ordinary'' Planck scale is simply $1/\sqrt{G} \sim 10^{19}$~GeV.
It is only an accident of history (Newton's law of gravity
predating general relativity, or for that matter Poisson's
equation) that it is defined this way, and the tradition is
continued by those with a great fondness for typing 
``{\tt 8$\backslash$pi}''.}
Planck scale $M_\P = 1/\sqrt{8\pi G} \sim 10^{18}$~GeV, so that 
we can (hopefully) ignore string theory (!).
We imagine an expanding universe with matter and radiation in
a thermal state, perfectly homogeneous and isotropic (we can
put in perturbations later), and all conserved quantum numbers
set to zero (no chemical potentials).  Note that asymptotic
freedom makes our task much easier; at the high temperatures
we are concerned with, QCD (and possible grand unified gauge
interactions) are weakly coupled, allowing us to work within
the framework of perturbation theory.

\subsection{Phase transitions}

The high temperatures and densities characteristic of the early
universe typically put matter fields into different phases than
they are in at zero temperature and density, and often these
phases are ones in which symmetries are restored
\cite{kt,lindebook}.  Consider a
simple theory of a real scalar field $\phi$ with a $\Z_2$
symmetry $\phi \rightarrow -\phi$.  The potential at zero
temperature might be of the form
\be  
  V(\phi, T=0) = -{{\mu^2}\over 2}\phi^2 + {\lambda\over 4}
  \phi^4\ .
  \label{phi4}
\ee
Interactions with a thermal background typically give 
positive contributions to the potential at finite temperature:
\be
  V(\phi, T) = V(\phi, 0) + \alpha T^2 \phi^2 + \cdots\ ,
\ee
where $\alpha = \lambda/8$ in the theory defined by (\ref{phi4}).

At high $T$, the coefficient of $\phi^2$ in the effective
potential, $(\alpha T^2 - {1\over 2}\mu^2)$, will be a 
positive number, so the minimum-energy state will be one 
with vanishing expectation value,
$\langle\phi\rangle = 0$.  The $\Z_2$ symmetry
is unbroken in such a state.  As the temperature declines,
eventually the coefficient will be negative and there
will be two lowest-energy states, with equal and opposite
values of $\langle\phi\rangle$.  A zero-temperature vacuum
will be built upon one of these values, which are not
invariant under the $\Z_2$; we therefore say the symmetry
is spontaneously broken.  The dynamics of the
transition from unbroken to broken symmetry is described
by a phase transition,
which might be either first-order or second-order.  
A first-order transition is one in which first derivatives of the
order parameter (in this case $\phi$) are discontinuous; they
are generally dramatic, with phases coexisting simultaneously,
and proceed by nucleation of bubbles of the new phase.  In a
second-order transition only second derivatives are discontinuous;
they are generally more gradual, without mixing of phases,
and proceed by ``spinodal decomposition''.  (I hope it is clear
that a huge amount of honest physics is being glossed over in
this brief discussion.)

\subsection{Topological defects}

Note that, post-transition, the field falls into the vacuum
manifold (the set of field values with minimum energy --- in
our current example it's simply two points) essentially
randomly.  It will fall in different directions at different
spatial locations $x_1$ and
$x_2$ separated by more than one correlation
length of the field.  In an ordinary FRW universe, the field
cannot be correlated on scales larger than approximately
$H^{-1}$, as this is the distance to the particle horizon
(as we will discuss below in the section on inflation).
If $\langle\phi(x_1)\rangle = +v$ and 
$\langle\phi(x_2)\rangle = -v$, then somewhere in between $x_1$ and
$x_2$ $\phi$ must climb over the energy barrier to
pass through zero.  Where this happens there will be energy
density; this is known as a ``topological defect'' (in this case
a defect of codimension one, a domain wall).  The argument that
the existence of horizons implies the production of defects
is known as the ``Kibble mechanism''.

More complicated
vacuum manifolds $\M$ will give other forms of defects, depending
on the topology of $\M$; if the homotopy group
$\pi_q(\M)$ (the set of topologically
inequivalent maps from $S^q$ into $\M$) is nontrivial, we will
have defects of (spatial) codimension $(q+1)$.  In three spatial
dimensions, nontrivial $\pi_0(\M)$ gives rise to walls (such as
in our example, for which $\pi_0(\M) = \Z_2$),  nontrivial 
$\pi_1(\M)$ gives rise to (cosmic) strings, and nontrivial
$\pi_2(\M)$ gives rise to pointlike defects (monopoles)
\cite{vilenkinshellard}.  Nobody will be upset if you refer to
these defects as ``branes''.

When a symmetry group $G$ is broken to a subgroup $H$, the vacuum
manifold is the quotient space $\M=G/H$, so we can determine
what sorts of defects might be created at an early-universe
phase transition.  A good tool for doing this is the exact
homotopy sequence \cite{steenrod,bcp}
\be
  \cdots \rightarrow  \pi_{q+1}(G/H) 
  \rightarrow
   \pi_{q}(H) \ \mapright{\alpha_q} \
   \pi_{q}(G) \ \mapright{\beta_q}\ \pi_{q}(G/H)\ 
   \mapright{\gamma_q} \ 
   \pi_{q-1}(H)  \rightarrow 
  \cdots  \rightarrow  \pi_0(G/H) \rightarrow  0\ ,
\ee
where $0$ is the trivial group.
The maps $\alpha_i$, $\beta_i$ and $\gamma_i$ are specified in terms
of the spaces $G$, $H$ and $G/H$, and they are all group
homomorphisms.  For example, the map $\alpha_q$ takes the image of a
$q$-sphere in $H$ into an image of a $q$-sphere in $G$ using the
inclusion of $H$ as a subgroup in $G$, $i:H\hookrightarrow G$.
``Exactness'' means that the image
of each map is precisely equal to the kernel (the set of elements
taken to zero) of the map following it.  

An important consequence of
exactness is that if two spaces $A$ and $B$ are sandwiched between the
trivial group, $0\rightarrow A\ \mapright\omega
\ B\rightarrow 0$, then the map $\omega$ must be an isomorphism.
This is easy to see: since the kernel of
$B\rightarrow 0$ is all of $B$, $\omega$ must be onto.  Meanwhile, since
the kernel of $\omega$ is the image of $0\rightarrow A$ (which is just
zero), in order for $\omega$ to be a group homomorphism it must be
one-to-one.  Thus, $\omega$ is an isomorphism.  You can also check
for yourself that the exact sequence $0 \rightarrow A \rightarrow
0$ implies that $A$ must be the trivial group.

The exact homotopy sequence can be used in conjunction with
our knowledge of various facts about the topology of Lie groups
to calculate $\pi_q(\M)$.  Some of the relevant facts include:
1.) For any Lie group $G$, $\pi_2(G)=0$.  2.) For any {\it simple}
group $G$, $\pi_3(G)=\Z$. 3.) $\pi_1(\su{n})=0$, but $\pi_1(\u{n})
=\Z$, and $\pi_1(\so{n>2}) = \Z_2$.  4.) $\pi_0$ simply counts
the number of disconnected pieces into which a space falls, so
$\pi_0(\su{n})=\pi_0(\so{n})=\pi_0(\u{n})=0$, and
$\pi_0(\o{n})=\Z_2$.  5.) Finally, for any spaces (not just
groups) $A$ and $B$, we have $\pi_q(A\times B) = \pi_q(A)
\times \pi_q(B)$.  For some examples of homotopy calculations
see \cite{bcp}.

As an example, consider $\su2$ breaking
down to $\u1$.  In the exact homotopy sequence, 
$\pi_0(\su2/\u1)$ and $\pi_1(\su2/\u1)$ are each sandwiched
between $0$'s, so both are trivial.  On the other hand, we
have
\be
  \pi_2(\su2)=0 \rightarrow \pi_2(\su2/\u1) \rightarrow
  \pi_1(\u1)={\bf Z} \rightarrow \pi_1(\su2)=0\ ,
\ee
so the map $\pi_2(\su2/\u1) \rightarrow {\bf Z}$ must be an
isomorphism, $\pi_2(\su2/\u1) = {\bf Z}$.  This theory
(the ``Georgi-Glashow model'') therefore predicts magnetic
monopoles with charges (proportional to the winding number of
the map $S^2 \rightarrow \su2/\u1$) taking values in $\Z$.
[The modifier ``magnetic'' is only appropriate if the original
$\su2$ was a gauge symmetry, in which case the monopole acts
as a source for the magnetic field of the unbroken $\u1$.
There can also be monopoles from the breakdown of a global
symmetry, although there are no solutions with finite energy.
Infinite energies aren't generally looked down upon by 
cosmologists, as the universe is a big place; more of a
worry would be an infinite energy density, which does not
occur in global monopoles.]

Once defects are produced at a phase transition, the question
of cosmological interest is how they subsequently evolve.  This
will be very different for different sorts of defects, and can
be altered by going beyond the simplest models \cite{vilenkinshellard}.
We will encounter some examples below.

\subsection{Relic particle abundances}
\label{subsec:relicabs}

One of the most useful things to do in cosmology is to calculate
the abundance of a given particle species from a specified initial
condition in the early universe.
First consider the properties of
particles in thermal equilibrium (with zero chemical potential).  
In the relativistic limit $m\ll T$, the number density $n$ and energy
density $\rho$ are given by
\bea
  n & \approx & T^3 \cr
  \rho & \approx & T^4 \ . \label{hotte}
\eea
Here we have begun what will be a conventional practice during
this lecture, ignoring factors of order unity.  To get them
right see any standard text \cite{peebles,kt}.  Note that the
Friedmann equation during a phase when the universe is flat
and radiation-dominated can be expressed simply as
\be
  H \approx {{T^2}\over {M_\P}}\ .
  \label{radhubble}
\ee
(The appearance of the Planck scale here isn't a sign of the
importance of quantum gravity, but merely classical gravity
plus the fact that we've set $\hbar = c= 1$.)
In the nonrelativistic limit ($m\gg T$), meanwhile, we have
\bea
  n & \approx & (mT)^{3/2} e^{-m/T} \cr
  \rho & \approx & mn\ . \label{coldte}
\eea
The energy density of nonrelativistic particles is just their
number density times the individual particle masses.

Particles will tend to stay in thermal equilibrium as
long as reaction rates $\Gamma$ are much faster than the
expansion rate $H$, so that the particles have plenty of
time to interact before the expansion of the universe
separates them.  A particle for which $\Gamma \ll H$ is
referred to as decoupled or ``frozen-out''; for species
which are kept in thermal equilibrium by the exchange of
massive bosons, $\Gamma \propto T^5$, and such particles
will be frozen-out at sufficiently low temperatures.
(Of course, a species may be noninteracting with the 
thermal bath and nevertheless in an essentially thermal
distribution, as we've already noted for the CMB; 
as another example, massless neutrinos decouple
while in a thermal distribution, which is then simply preserved
as the universe expands and the temperature decreases.)

There are two limiting cases of interest, decoupling while
relativistic (``hot relics'') and while nonrelativistic
(``cold relics'')\footnote{``Hot dark matter'', then, refers to
dark matter particles which were relativistic when they
decoupled --- not necessarily relativistic today.}.  
A hot relic $X$ will have a number density
at freeze-out approximately equal to the photon number
density,
\be
  n_X(T_f) \sim T_f^3 \sim n_\gamma(T_f)\ ,
\ee
where $T_f$ is the freeze-out temperature.  Subsequently,
the number densities of both $X$ and photons simply diminish
as the volume increases, $n_X\propto n_\gamma \propto a^{-3}$,
so their present-day number density is approximately
\be
  n_{X0} \sim n_{\gamma 0} \sim 10^2~{\rm cm}^{-3}\ .
\ee
We express this number as $10^2$ rather than $422$ since the
roughness of our estimate does not warrant
such misleading precision.  The leading correction to this value
is typically due to the production of additional photons
subsequent to the decoupling of $X$; in the Standard Model,
the number density of photons increases by a factor of
approximately $100$ between the electroweak phase transition
and today, and a species which decouples during this period
will be diluted by a factor of between $1$ and $100$ depending
on precisely when it freezes out.  So, for example, neutrinos 
which are light
($m_\nu < $~MeV) have a number density today of $n_\nu = 
115$~cm$^{-3}$ per species, and a corresponding contribution to the 
density parameter (if they are nevertheless heavy enough to be
nonrelativistic today) of
\be
  \Omega_{0, \nu} = \left({{m_\nu}\over{92~{\rm eV}}}\right) h^{-2}.
\ee
Thus, a neutrino with $m_\nu\sim 10^{-2}$~eV (as
might be a reasonable reading of the recent SuperKamiokande
data \cite{superk}) would contribute $\Omega_\nu\sim 2\times 10^{-4}$.
This is large enough to be interesting without being large enough
to make neutrinos be the dark matter.  That's good news, since
the large velocities of neutrinos make them free-stream out of
overdense regions, diminishing primordial perturbations and
leaving us with a universe which has much less structure on
small scales than we actually observe.  On the other hand, the
roughness of our estimates (and the data) leaves open the
possibility that neutrinos are nevertheless dynamically 
important, perhaps as part of a complicated mixture of dark
matter particles \cite{Primack:1998xd}.

For cold relics, the number density is plummeting rapidly during
freeze-out due to the exponential in (\ref{coldte}), and the
details of the interactions can be important.  But, very
roughly, the answer works out to be
\be
  n_X \sim {{n_\gamma}\over{\sigma_0 m_X M_\P}} \ ,
\ee
where $\sigma_0$ is the annihilation cross-section of $X$ at
$T=m_X$.  An example of a cold relic is provided by protons,
for which $m_p\sim 1$~GeV and $\sigma_0\sim m_\pi^{-2}
\sim (.1~{\rm GeV})^{-2}$.  This implies $n_p/n_\gamma 
\sim 10^{-20}$, which is rather at odds with the observed
value $n_p/n_\gamma \sim 10^{-10}$; this conflict brings home
the need for a sensible theory of baryogenesis
\cite{Dolgov:1992fr,Riotto:1998bt,Riotto:1999yt}.  (We might
worry that the disagreement between theory and observation 
in this case indicates that we had no clue how to really 
calculate relic abundances, if it weren't for the shining
counterexample of nucleosynthesis to be discussed below.)

A less depressing example of a cold relic is provided by weakly
interacting massive particles (``wimps''), a generic
name given to particles with cross-sections characteristic of
the weak interactions, $\sigma_0 \sim G_F \sim 
(300~{\rm GeV})^{-2}$.  Then the relic abundance today will be
\be
  n_{0, {\rm wimp}} \sim {{n_{\gamma 0}}\over{G_F m_{\rm wimp} M_\P}} 
  \sim \left({{10^{-13}~{\rm GeV}}\over{m_{\rm wimp}}}\right)
  ~{\rm cm}^{-3}\ ,  
\ee
which leads in turn to a density parameter
\be
  \Omega_{0, {\rm wimp}}\sim 1\ .\label{omegawimp}
\ee
The independence of (\ref{omegawimp}) on $m_{\rm wimp}$ (at 
least at our crude level of approximation) means that particles
with weak-interaction annihilation cross-sections provide 
excellent candidates for cold dark matter.  A standard example
is the lightest supersymmetric particle (``LSP'') 
\cite{Kolb:1998kj,Ellis:1998gt}.

\subsection{Vacuum displacement}
\label{subsec:vacdis}

Another important possibility is the existence of relics
which were never in thermal equilibrium.  An example of these has
already been discussed:  the production of topological defects
at phase transitions.  Let's discuss another kind of 
non-thermal relic, which derives from what we might call
``vacuum displacement''.  Consider the action for a real scalar
field in curved spacetime (assumed to be four-dimensional):
\be
  S = \int d^4x\, \sqrt{-g}\left[-{1\over 2}g^{\mu\nu}
  \partial_\mu\phi \partial_\nu\phi - V(\phi)\right]\ .
\ee
If we assume that $\phi$ is spatially homogeneous ($\partial_i\phi
= 0$), its equation of motion in
the Robertson-Walker metric (\ref{rwmetric}) will be
\be
  \ddot\phi + 3H\dot\phi + V'(\phi) = 0\ ,\label{phifrw}
\ee
where an overdot indicates a partial derivative with respect
to time, and a prime indicates a derivative with respect to $\phi$.
For a free massive scalar field, $V(\phi) = {1\over 2} m_\phi^2
\phi^2$, and (\ref{phifrw}) describes a harmonic oscillator with
a time-dependent damping term.  For $H > m_\phi$ the
field will be overdamped, and stay essentially constant at 
whatever point in the potential it finds itself.  So let us
imagine that at some time in the very early universe (when $H$
was large) we had such an overdamped homogeneous scalar field,
stuck at a value $\phi = \phi_*$; the total energy density in the field 
is simply the potential energy ${1\over 2}m_\phi^2\phi_*^2$.  
The Hubble parameter $H$ will decrease to approximately $m_\phi$ 
when the temperature reaches $T_*=\sqrt{m_\phi M_\P}$, 
after which the field
will be able to evolve and will begin to oscillate in its
potential.  The vacuum energy is converted to a combination of
vacuum and kinetic energy which will redshift like matter,
as $\rho_\phi\propto a^{-3}$; in a particle interpretation, the
field is a Bose condensate of zero-momentum particles.  We
will therefore have
\be
  \rho_\phi(a) \sim {1\over 2}m_\phi^2 \phi_*^2\left(
  {{a_*}\over a}\right)^3\ ,
\ee
which leads to a density parameter today
\be
  \Omega_{0, \phi} \sim \left({{\phi_*^4 m_\phi}\over{10^{-19}
  ~{\rm GeV}^5}}\right)^{1/2}\ .
  \label{omegadisp}
\ee

A classic example of a non-thermal relic produced by vacuum 
displacement is the QCD axion, which has a typical primordial value
$\langle\phi\rangle\sim f_{\rm PQ}$ and a mass $m_\phi \sim
\Lambda_{\rm QCD}^2/f_{\rm PQ}$, where $f_{\rm PQ}$ is the
Peccei-Quinn symmetry-breaking scale and $\Lambda_{\rm QCD}\sim
0.3$~GeV is the QCD scale \cite{kt}.  In this case, plugging
in numbers reveals
\be
  \Omega_{0, \phi} \sim \left({{f_{\rm PQ}}\over{10^{13}~{\rm GeV}}}
  \right)^{3/2}\ .
  \label{vacdisomega}
\ee
The Peccei-Quinn scale is essentially a free parameter from a
theoretical point of view, but experiments and astrophysical
constraints have ruled out most values except for a small window
around $f_{\rm PQ}\sim 10^{12}~{\rm GeV}$.  The axion therefore
remains a viable dark matter candidate 
\cite{Kolb:1998kj,Ellis:1998gt}.  Note that, even though 
dark matter axions are very light
($\Lambda_{\rm QCD}^2/f_{\rm PQ}\sim 10^{-4}$~eV), they are
extremely non-relativistic, which can be traced to the non-thermal
nature of their production process.  (Another important way to
produce axions is through the decay of axion cosmic strings
\cite{kt,vilenkinshellard}.)

\subsection{Thermal history of the universe}
\label{subsec:thermhist}

We are now empowered to take a brief tour through the evolution
of the universe, starting at a temperature $T\sim 10^{16}$~GeV,
and assuming the correctness of the Standard Model plus perhaps
some grand unified theory, but nothing truly exotic.  (At temperatures
higher than this, not only do we have to worry about quantum
gravity, but the Hubble parameter is so large that essentially
no perturbative interactions are able to maintain thermal
equilibrium; either strong interactions are important, or 
every species is frozen out.)  The first
event we encounter as the universe expands is the grand unification
phase transition (if there is one).  
Here, some grand unified group $G$ breaks to 
the standard model group\footnote{You will often hear it said that
the standard model gauge group is $\su3\times\su2\times\u1$, but
this is not strictly correct; there is a $\Z_6$ subgroup leaving
all of the standard-model fields invariant.  The Lie algebras of
the two groups are identical, which is usually all that particle
physicists care about, but when topology is important it is safer
to keep track of the global structure of the group.}
$[\su3\times\su2\times\u1]/\Z_6$, with 
popular choices for $G$ including $\su5$, $\so{10}$, and E$_6$.

Most interesting particles decay away after the GUT transition,
with the possible exception of the all-important baryon asymmetry.
As Sakharov long ago figured out, to make a baryon asymmetry we need
three conditions \cite{Dolgov:1992fr,Riotto:1998bt,Riotto:1999yt}:
\begin{enumerate}
  \item Baryon number violation.
  \item $C$ and $CP$ violation.
  \item Departure from thermal equilibrium.
\end{enumerate}
The $X$-bosons of GUTs typically have decays which can violate
$B$, $C$, and $CP$.  Departure from equilibrium happens because
the $X$'s first freeze out, then decay.  With the right choice of
parameters, we can get $n_{\rm B}/n_\gamma\sim 10^{-10}$, the
sought-after number.

One problem with this scenario is that, 
at $T > T_{\rm EW}$, nonperturbative effects in
the standard model (sphalerons) can violate baryon number.  These
will tend to restore the baryon number to its equilibrium value
(zero).  A potential escape is to notice that sphalerons violate
$B$ and lepton number $L$ but preserve the combination $B-L$,
so that for every excess baryon produced a corresponding lepton
must be produced.  If our GUT generates a nonzero $B-L$ it will
therefore survive, as it cannot be changed by standard model
processes.  The $\su5$ theory conserves $B-L$ and is therefore
apparently not the origin of the baryon asymmetry, although
$B-L$ can be generated in $\so{10}$ models
\cite{Riotto:1998bt,Riotto:1999yt}.

Another worry about GUTs is the prediction of
magnetic monopoles. Since
$\pi_1(G)=0$ for any simple Lie group $G$, we have
$\pi_2(G/H)=\pi_1([\su3\times\su2\times\u1]/\Z_6)=\Z$, and monopoles
are inescapable.  We end up with
\be
  \Omega_{0, {\rm mono}} \sim 10^{11} \left({{T_{\rm GUT}}\over
  {10^{14}~{\rm GeV}}}\right)^3 \left({{m_{\rm mono}}\over
  {10^{16}~{\rm GeV}}}\right)\ .
\ee
This is far too big; the monopole abundance in GUTs is a
serious problem, one which can be
solved by inflation (which we will discuss later).

Depending on the details of the symmetry group being broken,
the GUT phase transition can also produce domain walls (which
also disastrously overdominate the universe) or cosmic strings
(which will not dominate the energy density, and in fact may
have various beneficial effects) \cite{vilenkinshellard}.

Below the GUT temperature, nothing really happens (as far
as we know) until $T_{\rm EW}\sim 300$~GeV ($z\sim 10^{15}$), when
the Standard Model gauge symmetry $[\su3\times\su2\times\u1]/\Z_6$
is broken to $\su3\times\u1$.
No topological defects are produced.  (Magnetic fields may be,
however; see for example 
\cite{Baym:1996fk,Joyce:1997uy,Tornkvist:1998hd}.)  
Most interestingly, the electroweak phase transition may be
responsible for baryogenesis 
\cite{Dolgov:1992fr,Riotto:1998bt,Riotto:1999yt}.
The nonperturbative $B$-violating interactions of the Standard
Model are exponentially suppressed after the phase transition, so
any asymmetry generated at that time will be preserved.
The important question is the amount of $CP$ violation and 
departure from thermal equilibrium.  Both exist in the minimal
Standard Model; $CP$ violation is present in the CKM matrix,
and expansion of the universe provides some departure from
equilibrium.  Both are very small, however; the amounts are
apparently not nearly enough to generate the required 
asymmetry.

It is therefore necessary to augment the Standard Model.
Fortunately the simple action of adding additional Higgs
bosons can work both to increase the amount of $CP$ 
violation (by introducing new mixing angles) and the
departure from thermal equilibrium (by changing the phase
transition from second order, which it is in the 
SM for experimentally allowed values of the Higgs mass,
to first order).  Supersymmetric extensions of the SM
require an extra Higgs doublet in addition to the one of
the minimal SM, so there is some hope for a \susy
scenario.  At this point, however, the relevant dynamics
at the phase transition are not sufficiently well understood
for us to say whether electroweak baryogenesis is a
sensible idea.  (It does, however, have the pleasant
aspect of being related to experimentally testable aspects
of particle physics.)

There is one more scenario worth mentioning, known as 
Affleck-Dine baryogenesis
\cite{Affleck:1985fy,Dolgov:1992fr}.  The idea here is to have a scalar
condensate with energy density produced by vacuum displacement,
but to have the scalar carry baryon number.  Its decay can
then lead to the observed baryon asymmetry.

After the electroweak transition, the next interesting
event is the QCD phase transition at $T_{\rm QCD}\sim
0.3$~GeV.  Actually there are two things that happen, lumped
together for convenience as the ``QCD phase transition'':
chiral symmetry breaking, and the confinement of quarks and
gluons into hadrons.  Our understanding of
the QCD transition is also underdeveloped, although it is
likely to be second order and does not lead to any important
relics \cite{Smilga:1998ki}.

At a temperature of $T_f\sim 1$~MeV, the weak interactions
freeze out, and free neutrons
and protons decouple.  The neutron to proton ratio at this time
is approximately $1/6$, and gradually decreases as the neutrons
decay.  Soon thereafter, at around
$T_{\rm BBN}\sim 80$~keV, almost all of the neutrons fuse
with protons into light elements (D, $^3$He, $^4$He, Li), 
a process known as ``Big Bang Nucleosynthesis'' 
\cite{schrammturner,mendozahogan,sarkar96}.  Although it would
seem to be a rather mundane low-energy phenomenon from the
lofty point of view of constructing a theory of everything,
the results of BBN are actually of great importance to string
theory (or any other theories which could affect cosmology),
since they offer by far the best empirical
constraints on the behavior of the universe at relatively
early times.

The abundances of light elements, like those of any other
relics, depend on the interplay between interaction rates
$\Gamma_i$ of species $i$ and the Hubble parameter $H$.  The
reaction rates depend in turn on the baryon to photon ratio
$n_{\rm B}/n_\gamma$, not to mention the parameters of the 
Standard Model (the fine-structure constant $\alpha$, the 
Fermi constant $G_{\rm F}$, the electron mass $m_e$, etc.).
Since BBN occurs well into the radiation-dominated era, the 
expansion rate is 
\be
  H^2 = {1\over 3M_\P^2}\rho_{\rm R}\ .
  \label{bbnfeq}
\ee
In the standard picture, $\rho_{\rm R}$ comes essentially from
photons (whose density we can count) and neutrinos (whose density
per species we have calculated above), as well as electrons when
$T > m_e$.

It is a remarkable fact that the observed light-element
abundances, coupled with the observed number of light neutrino
species $N_\nu = 3$, are consistent with the BBN prediction for
$n_{\rm B}/n_\gamma \approx 5\times 10^{-10}$, a number which is 
consistent with the observed ratio of baryons to photons.
(Consistent in the sense of being not incompatible; in fact
the observed number of baryons is somewhat lower, but there's
nothing stopping some of the baryons from being dark
\cite{Fukugita:1997bi}.)  The agreement, furthermore, 
is not with a single number, but the individual abundances of 
D, $^4$He, and $^7$Li.  Not only does this give us confidence
in our ability to calculate relic abundances (both of nuclei
and of the neutrinos that enter the calculation), it also
implies that the current values of $n_{\rm B}/n_\gamma$,
the number density of hot relics,
Newton's constant $G$, the fine structure constant
$\alpha$, and all of the other parameters of physics that
enter the calculation, are similar to what their values were
at the time of BBN, when the universe was only 1 second old
\cite{Campbell:1995bf}.
This is astonishing when we consider the number of ways in
which they could have varied, as discussed briefly in the next
section \cite{Barrow:1997qh}.

Apart from constraints on specific models, nucleosynthesis also 
provides the best evidence that the early universe was in a hot
thermal state, with dynamics governed by the conventional 
Friedmann equation.  Although it is possible to imagine alternative
early histories which are compatible with the observed 
light-element abundances, it would be surprising if any
dramatically different model led coincidentally to the same
predictions as the conventional picture.

\subsection{Gravitinos and moduli}
\label{subsec:gravmod}

An example of a model constrained by BBN 
is provided by any theory of supergravity in which
\SUSY is broken at an intermediate scale $M_{\rm I}\sim 10^{11}$~GeV
in a hidden sector (the gravitationally mediated models).
In these theories the gravitino, the superpartner of the graviton,
will have a mass
\be
  m_{3/2} \sim M_{\rm I}^2/M_\P \sim 10^3~{\rm GeV}
\ee
(which is also the scale of \SUSY breaking in the visible
sector).  The gravitino is of special interest since its
interactions are so weak (its couplings, gravitational in origin,
are suppressed by powers of $M_\P$)
implying that 1.) it decouples early, while relativistic,
leaving a large relic abundance, and 2.) it decays slowly
and therefore relatively late.  Indeed, the lifetime is
\be
  \tau_{3/2} \sim M_\P^2/m_{3/2}^3 \sim 10^{27}~{\rm GeV}^{-1}
  \sim 10^3~{\rm sec}\ ,
\ee
somewhat after nucleosynthesis.  The decaying
gravitinos produce a large number of high-energy photons,
which can both dilute the baryons and photodissociate the
nuclei, changing their abundances and thereby ruining the
agreement with observation.  This ``gravitino problem'' might
be alleviated by inflation (as we will later discuss), but
serves as an important constraint on specific models 
\cite{Moroi:1995fs,Kallosh:2000jj,gtr,grt,Buonanno:2000cp}.

The success of BBN also places limits on the time variation
of the coupling constants of the Standard Model\footnote{Note
that there are a number of other constraints
on such time dependence, including solar-system tests of
gravity, the relative spacing of absorption lines in quasar
spectra, and isotopic abundances in the Oklo natural
reactor \cite{damour,oklo}.}.  In string theory,
these couplings are all related to the expectation values of
moduli (scalar fields parameterizing ``flat directions'' in
field space which arise due to the constraints of supersymmetry),
and could in principle vary with time \cite{Maeda:1988ku}.  The fact
that they don't vary is most easily accommodated by imagining 
that the moduli are sitting at the minima of some potentials;
in fact this is completely sensible given that supersymmetry
is broken, so we expect that $m_{\rm moduli}\sim M_{\rm SUSY}
\sim 10^3$~GeV, enough to fix their values for all temperatures
less than $T\sim \sqrt{M_{\rm SUSY}M_\P}\sim 10^{11}$~GeV (although
at higher temperatures they could vary in interesting ways).

On the other hand, massive moduli present their own problems.
They are produced as non-thermal relics due to vacuum 
displacement \cite{Coughlan:1983ci,deCarlos:1993jw,Banks:1994en}.  
At high temperatures the fields
are at some random point in moduli space, which will typically
be of order $\phi_*\sim M_\P$.  If the moduli were stable,  from
(\ref{vacdisomega}) we
would therefore expect a contribution to the critical density
of order
\be
  \Omega_{0, {\rm moduli}} \sim 10^{27}\ .
\ee
This number is clearly embarrassingly big, and something has to
be done about it.  The moduli can of course decay into other
particles, but their lifetimes are similar to those of gravitinos,
and their decay also tends to destroy the success of BBN.  Due
to their different production mechanism, it is harder to dilute
the moduli abundance during inflation (since the scalar vev can
remain displaced while inflation occurs), and the ``moduli
problem'' poses a significant puzzle for string theories.
(One promising solution would be the existence of a
point of enhanced symmetry which would make the high-temperature
and low-temperature minima of the potential coincide 
\cite{Dine:1995uk}.)

In addition to the overproduction of moduli, there are also
problems with their stabilization, especially for the dilaton,
perhaps the best-understood example of a modulus field.
One problem is that the dilaton expectation value acts as a
coupling constant in string theory, and very general arguments
indicate that the dilaton cannot be stabilized at a value we would
characterize as corresponding to weak coupling \cite{Dine:1985he}.
Another is that, in certain popular models for stabilizing
the dilaton using gaugino condensates, the cosmological evolution
would almost inevitably tend to overshoot the desired minumum
of the dilaton potential and run off to an anti-de~Sitter
vacuum \cite{Brustein:1993nk}.  Problems such as these are the
subject of current investigation \cite{Huey:2000jx,Dine:2000ds}.

There are numerous aspects of the cosmology of moduli
which can't be covered here; see Michael Dine's TASI lectures
for an overview \cite{Dine:2000bf}.

\subsection{Density fluctuations}

The subject of primordial density fluctuations and their evolution
into galaxies is a huge subject in its own right
\cite{Liddle:1993fq,dodelsongatesturner,bahcall,primack}, 
which time and space did not permit covering in these
lectures.  By way of executive summary, models in which
the matter density is dominated by cold dark matter (CDM) and the
perturbations are nearly scale-free, adiabatic, and Gaussian
(just as predicted by inflation --- see section 
(\ref{subsec:perts}) below)
are relatively good fits to the data.  Such models are often
compared to the fiducial $\Omega_{\rm M} = 1$ case
(``Standard CDM''), which cannot simultaneously be fit to
the CMB anisotropy amplitude and the amount of structure seen
in redshift surveys.  Since it is harder to change the CMB
normalization, modifications of the CDM scenario need to
decrease the power on small scales in order to fit the
galaxy data.  Fortunately, most such modifications ---
a nonzero cosmological constant (``$\Lambda$CDM''), an open 
universe (``OCDM''), an admixture of hot dark matter such as
neutrinos (``$\nu$CDM'') --- work in this direction.  The most
favored model at the moment is that with an appreciable
cosmological constant, although none of the models is perfect.

Hot dark matter models are completely ruled out if they are
based on scale-free adiabatic perturbation spectra.  There is
also the possibility of seeding perturbations with ``seeds''
such as topological defects, although such scenarios are
currently disfavored for their failure to fit the CMB 
anisotropy spectrum (for examples of recent analyses
see \cite{Sakellariadou:1999ru,Albrecht:2000hu,Pogosian:2000pn}.)

\section{Inflation}
\label{sec:inflation}

\subsection{The idea}

Despite the great success of the
conventional cosmology, there remain two interesting
conceptual puzzles: flatness and isotropy.  The leading
solution to these problems is the inflationary universe
scenario, which has become a central organizing principle
of modern cosmology \cite{Linde:1994yf,Kolb:1996rc,Turner:1997we,
brand97,unruh,Liddle:1999mq,Guth:2000ka}.

The flatness problem comes from considering the Friedmann
equation in a universe with matter and radiation but no vacuum
energy:
\be
  H^2 = {1\over{3M_\P^3}}(\rho_{\rm M} + \rho_{\rm R})
  -{k\over{a^2}}\ .
\ee
The curvature term $-k/a^2$ is proportional to $a^{-2}$
(obviously), while the energy density terms fall off faster
with increasing scale factor, $\rho_{\rm M}\propto a^{-3}$ and
$\rho_{\rm R}\propto a^{-4}$.  This raises the question of why
the ratio $(ka^{-2})/(\rho/3M_p^2)$ isn't much larger than
unity, given that $a$ has increased by a factor of perhaps
$10^{28}$ since the grand unification epoch.  Said another
way, the point $\Omega = 1$ is a repulsive fixed point ---
any deviation from this value will grow with time, so why
do we observe $\Omega\sim 1$ today?

The isotropy problem is also called the ``horizon problem'',
since it stems from the existence of particle horizons in
FRW cosmologies.  Horizons exist because there is only a
finite amount of time since the Big Bang singularity, and
thus only a finite distance that photons can travel within
the age of the universe.  Consider a photon moving along
a radial trajectory in a flat universe (the
generalization to nonflat universes is straightforward).
A radial null path obeys
\be
  0 = ds^2 = -dt^2 + a^2 dr^2\ ,
\ee
so the comoving distance traveled by such a photon between
times $t_1$ and $t_2$ is
\be
  \Delta r = \int^{t_2}_{t_1} {{dt}\over{a(t)}}\ .
\ee
(To get the physical distance as it would be measured by an
observer at time $t_1$, simply multiply by $a(t_1)$.)
For a universe dominated by an energy density $\rho\propto a^{-n}$,
this becomes
\be
  \Delta r = {1\over{a_*^{n/2}H_*}}\left({2\over{n-2}}\right)
  \Delta(a^{n/2-1})  \ ,
\ee
where the $*$ subscripts refer to some fiducial epoch (the
quantity $a_*^{n/2}H_*$ is a constant).  The horizon
problem is simply the fact that the CMB is isotropic to a high
degree of precision, even though widely separated points on the
last scattering surface are completely outside each others'
horizons.  Choosing $a_0=1$, the comoving horizon size today
is approximately $H_0^{-1}$, which is also the approximate
comoving distance between us and the surface of last scattering
(since, of the comoving distance traversed by a photon between
a redshift of infinity and a redshift of zero, the amount
between $z=\infty$ and $z=1100$ is much less than the amount
between $z=1100$ and $z=0$).  Meanwhile, the comoving horizon size
at the time of last scattering was approximately $a_{\rm CMB}H_0^{-1}
\sim 10^{-3}H_0^{-1}$, so distinct patches of the CMB sky were
causally disconnected at recombination.  Nevertheless, they
are observed to be at the same temperature to high precision.
The question then is, how did they know ahead of time to
coordinate their evolution in the right way, even though they
were never in causal contact?  We must somehow modify the
causal structure of the conventional FRW cosmology.

Now let's consider modifying the conventional picture by 
positing a period in the early universe when it was dominated
by vacuum energy rather than by matter or radiation.  (We will
still work in the context of a Robertson-Walker metric, which
of course assumes isotropy from the start, but we'll come back
to that point later.)  Then the flatness and horizon problems
can be simultaneously solved.  First, during the vacuum-dominated
era, $\rho/3M_p^2 \propto a^0$ grows rapidly with respect to
$-k/a^2$, so the universe becomes flatter with time ($\Omega$ is
driven to unity).  If this process proceeds for a sufficiently
long period, after which the vacuum energy is converted into
matter and radiation, the density parameter will be sufficiently
close to unity that it will not have had a chance to noticeably
change into the present era.  The horizon problem, meanwhile,
can be traced to the fact that the physical distance between
any two comoving objects grows as the scale factor, while the
physical horizon size in a matter- or radiation-dominated
universe grows more slowly, as $r_{\rm hor} \sim a^{n/2 - 1}H_0^{-1}$.
This can again be solved by an early period of exponential
expansion, in which the true horizon size grows to a fantastic
amount, so that our horizon today is actually much larger
than the naive estimate that it is equal to the Hubble
radius $H_0^{-1}$.

In fact, a truly exponential expansion is not necessary; both
problems can be solved by a universe which is accelerated,
$\ddot a > 0$.  Typically we require that this accelerated
period be sustained for 60 or more $e$-folds, which is what
is needed to solve the horizon problem.  It is easy to overshoot,
and this much inflation generally makes the present-day
universe spatially flat to incredible precision.

\subsection{Implementation}

Now let's consider how we can get an inflationary phase in the
early universe.  The most straightforward way is to use the
vacuum energy provided by the potential of a scalar field
(called the ``inflaton'').
Imagine a universe dominated by the energy of a spatially
homogeneous scalar.  The equations of motion include
(\ref{phifrw}), the equation of motion for a scalar
field in an RW metric:
\be
  \ddot\phi + 3H\dot\phi + V'(\phi) = 0\ ,\label{phifrw2}
\ee
as well as the Friedmann equation:
\be
  H^2 = {1\over{3M_\P^2}}\left({1\over 2}\dot\phi^2
  + V(\phi)\right)\ .
\ee
We've ignored the curvature term, since inflation will flatten
the universe anyway.  Inflation can occur if the evolution of
the field is sufficiently gradual that the potential energy
dominates the kinetic energy, and the second derivative of $\phi$
is small enough to allow this state of affairs to be maintained
for a sufficient period.  Thus, we want
\bea
  \dot\phi^2 & \ll & V(\phi)\ ,\cr
  |\ddot \phi | & \ll & |3H\dot\phi|,\ |V'|\ .
\eea
Satisfying these conditions requires the smallness of two 
dimensionless quantities known as ``slow-roll parameters'':
\bea
  \epsilon & = & {1\over 2}M_\P^2\left({{V'}\over V}\right)^2\ ,\cr
  \eta & = & M_\P^2\left({{V''}\over V}\right)\ .
  \label{srp}
\eea
(Note that $\epsilon \geq 0$, while $\eta$ can have either sign.
Note also that these definitions are not universal; some people
like to define them in terms of the Hubble parameter rather than
the potential.)
When both of these quantities are small we can have a prolonged
inflationary phase.  They are not sufficient, however; no matter
what the potential looks like, we can always choose initial conditions
with $|\dot\phi|$ so large that slow-roll is never applicable.
However, ``most'' initial conditions are attracted to an inflationary
phase if the slow-roll parameters are small.

It isn't hard to invent potentials which satisfy the slow-roll
conditions.  Consider perhaps the simplest possible example,
$V(\phi) = {1\over 2}m^2\phi^2$ (following the example in
\cite{Liddle:1999mq}).  In this case
\be
  \epsilon = \eta = {{2M_\P^2}\over{\phi^2}}\ .
\ee
Clearly, for large enough $\phi$, we can get the slow-roll
parameters to be as small as we like.  However, we have the
constraint that the energy density should not be as high as
the Planck scale, so that our classical analysis makes sense;
this implies $\phi \ll M_\P^2/m$.  If we start the field at a
value $\phi_i$, the number of $e$-folds before inflation ends
({\it i.e.}, before the slow-roll parameters become of order
unity) will be
\bea
  N &=& \int^{t_e}_{t_i} H\, dt \cr
  &\approx & -M_\P^{-2}\int^{\phi_e}_{\phi_i}{V\over{V'}}\, d\phi \cr
  & \approx & {{\phi_i^2}\over{4M_p^2}} - {1\over 2}\ .
\eea
The first equality is always true, the second uses the slow-roll
approximation, and the third is the result for this particular model.
To get 60 $e$-folds we therefore need $\phi_i > 16M_\P$.  Together
with the upper limit on the energy density, we find that there
is an upper limit on the mass parameter, $m \ll M_\P/16$.  In
fact the size of the observed density fluctuations puts a more
stringent upper limit on $m$, as we will discuss below.  But there
is no lower limit on $m$, so it is easy to obtain appropriate
inflationary potentials if only we are willing to posit large
hierarchies $m \ll M_\P$, or equivalently a small dimensionless
number $m/M_\P$.  Going through the same exercise with a $\lambda
\phi^4$ potential would have yielded a similar conclusion, that
$\lambda$ would have had to be quite small; we often say that the
inflaton must be weakly coupled.
(Of course, there is a sense in which
we are cheating, since for field values $\phi > M_\P$ we should
expect nonrenormalizable terms in the effective potential, of
the form $M_\P^{4-n}\phi^n$ with $n > 4$, to become important.
So in a realistic model it can be quite hard to get an
appropriate potential.)

At some point inflation ends, and the energy in the inflaton
potential is converted into a thermalized gas of matter and
radiation, a process known as ``reheating''.  
It used to be modeled as a perturbative decay of $\phi$-bosons
into other particles; this is a relatively inefficient process,
and the temperature of the resulting thermal state cannot be
very high.  More recently it has been realized that nonlinear
effects (parametric resonance) can efficiently transfer energy
from coherent oscillations of $\phi$ into other particles, a 
process referred to as ``preheating'' \cite{kls,bcdhk}.  
The resulting temperature
can be quite a bit higher than had been previously believed.
(On the other hand, we have already noted that the inflaton
tends to be weakly coupled, which suppresses the reheat 
temperature.)

A proper understanding of the reheating process is of utmost
importance, as it controls the production of various relics
that we may or may not want in our universe.  For example,
one of the most beneficial aspects of inflation in the context
of grand unification is that it can solve the monopole problem.
Essentially, any monopoles will be inflated away, leaving a
relic abundance well under the observational limits.  It is
therefore important that reheating does not reproduce too 
many monopoles (it almost certainly doesn't).  On the other
hand, we do want to reheat to a sufficiently high temperature
to allow for some sort of baryogenesis scenario.

It is nevertheless important to try to implement inflation
within a believable particle physics model, although we only
have time to telegraphically list some relevant issues.  

\begin{itemize}

\item A great deal of 
effort has gone into exploring the relationship between inflation
and supersymmetry, although simultaneously satisfying the
strict requirements of inflation and \SUSY turns out to be a
difficult task \cite{Randall:1997kx,Lyth:1998xn}.

\item Hybrid inflation is a kind of model which invokes two scalar
fields with a ``waterfall'' potential \cite{Linde:1994cn,Copeland:1994vg}.
One field rolls slowly and is weakly coupled, the other 
is strongly coupled and leads to efficient reheating once
the first rolls far enough. 

\item Another interesting class of models involve
scalar-tensor theories and make intimate use of the 
conformal transformations relating these theories to conventional
Einstein gravity \cite{Wands:1994uu}.

\item The need for a flat potential for the inflaton, coupled 
with the fact that string theory moduli can naturally have flat
potentials, makes the idea of ``modular inflation'' an
attractive one \cite{Banks:1995dp,Banks:1999dh}.  Specific
implementations have been studied, but we probably don't 
understand enough about moduli at this point to be confident of
finding a compelling model.

\end{itemize}

\subsection{Perturbations}
\label{subsec:perts}

A crucial element of inflationary scenarios is the production
of density perturbations, which may be the origin of the CMB
temperature anisotropies and the large-scale structure in 
galaxies that we observe today.  

The idea behind density perturbations generated by inflation
is fairly straightforward (it is only the conventions that
are a headache; look in the references to get numerical factors right
\cite{peebles,kt,Mukhanov:1992me,Liddle:1993fq,Lidsey:1995np,
Turner:1997we,Lyth:1998xn}).  
Inflation will attenuate any ambient particle density rapidly
to zero, leaving behind only the vacuum.  But the vacuum state
in an accelerating universe has a nonzero temperature, the
Gibbons-Hawking temperature, analogous to the Hawking temperature
of a black hole.  For a universe dominated by a potential energy
$V$ it is given by
\be
  T_{\rm GH} = H/2\pi \sim V^{1/2}/ M_P\ .
\ee
Corresponding to this temperature are fluctuations in the inflaton
field $\phi$ at each wavenumber $k$, with magnitude
\be
  |\Delta\phi|_k = T_{\rm GH}\ .
\ee
Since the potential is by hypothesis nearly flat, the fluctuations
in $\phi$ lead to small fluctuations in the energy density,
\be
  \delta\rho = V'(\phi) \delta\phi\ .
\ee
Inflation therefore produces density perturbations on every scale.
The amplitude of the perturbations is nearly equal at
each wavenumber, but there will be slight deviations due to the
gradual change in $V$ as the inflaton rolls.  We can
characterize the fluctuations in terms of their spectrum
$A_{\rm S}(k)$, related to the potential via
\be
  A_{\rm S}^2(k) \sim \left.{{V^3}\over{M_\P^6(V')^2}}\right|_{k=aH}\ ,
  \label{scalarspectrum}
\ee
where $k=aH$ indicates that the quantity $V^3/(V')^2$ is to be
evaluated at the moment when the physical scale of the
perturbation $\lambda=a/k$ is equal to the Hubble radius
$H^{-1}$.  Note that the actual normalization of 
(\ref{scalarspectrum}) is convention-dependent, and should drop
out of any physical answer.

The spectrum is given the subscript ``S'' because it describes
scalar fluctuations in the metric.  These are tied to the
energy-momentum distribution, and the density fluctuations
produced by inflation are adiabatic (or, better, ``isentropic'')
--- fluctuations in the density of all species
are correlated.  The fluctuations are also Gaussian, in the
sense that the phases of the Fourier modes describing fluctuations
at different scales are uncorrelated.  These aspects of 
inflationary perturbations --- a nearly scale-free spectrum
of adiabatic density fluctuations with a Gaussian distribution ---
are all consistent with current observations of the CMB and 
large-scale structure, and new data scheduled to be collected
over the next decade should greatly improve the precision of 
these tests.

It is not only the nearly-massless inflaton that is excited
during inflation, but any nearly-massless particle.  The
other important example is the graviton, which corresponds
to tensor perturbations in the metric (propagating excitations
of the gravitational field).  Tensor fluctuations have a spectrum
\be
  A_{\rm T}^2(k) \sim \left.{{V}\over{M_\P^4}}\right|_{k=aH}\ .
\ee
The existence of tensor perturbations is a crucial prediction
of inflation which may in principle be verifiable through
observations of the polarization of the CMB.  In practice,
however, the induced polarization is very small, and we may never
detect the tensor fluctuations even if they are there.

For purposes of understanding observations, it is useful to
parameterize the perturbation spectra in terms of observable
quantities.  We therefore write
\be
  A_{\rm S}^2(k) \propto k^{n_{\rm S}-1}
\ee
and
\be
  A_{\rm T}^2(k) \propto k^{n_{\rm T}}\ ,
\ee
where $n_{\rm S}$ and $n_{\rm T}$ are the ``spectral indices''.  They are
related to the slow-roll parameters of the potential by
\be
  n_{\rm S} = 1 -6\epsilon + 2\eta
\ee
and
\be
  n_{\rm T} = -2\epsilon\ .
\ee
Since the spectral indices are in principle observable, we
can hope through relations such as these to glean some information
about the inflaton potential itself.

Our current knowledge of the amplitude of the perturbations
already gives us important information about the energy scale
of inflation.  Note that the tensor perturbations depend on
$V$ alone (not its derivatives), so observations of tensor
modes yields direct knowledge of the energy scale.
If the CMB anisotropies seen by COBE are due to tensor
fluctuations (possible, although unlikely), we can instantly
derive $V_{\rm inflation}\sim (10^{16}$~GeV$)^4$.  
(Here, the value of $V$ being constrained is that which was
responsible for creating the observed fluctuations; namely,
60 $e$-folds before the end of inflation.)  This is 
remarkably reminiscent of the grand unification scale, which
is very encouraging.  Even in the more likely case that the
perturbations observed in the CMB are scalar in nature, we
can still write
\be
  V_{\rm inflation}^{1/4}\sim \epsilon^{1/4}10^{16}~{\rm GeV}\ ,
\ee
where $\epsilon$ is the slow-roll parameter defined in (\ref{srp}).
Although we expect $\epsilon$ to be small, the $1/4$ in the
exponent means that the dependence on $\epsilon$ is quite weak;
unless this parameter is extraordinarily tiny, it is very
likely that $V_{\rm inflation}^{1/4}\sim 10^{15}$-$10^{16}$~GeV.  The
fact that we can have such information about such tremendous
energy scales is a cause for great wonder.  

\subsection{Initial conditions and eternal inflation}

We don't have time to do justice to the interesting
topic of initial conditions for inflation.  It is an especially
acute subject once we realize that, although inflation is
supposed to solve the horizon problem, it is necessary to
start the universe simultaneously inflating in a region
larger than one horizon volume in order to achieve successful
inflation \cite{Vachaspati:1998dy}.  Presumably we must
appeal to some sort of quantum fluctuation to get the universe
(or some patch thereof) into such a state.

Fortunately, inflation has the wonderful property that it
is eternal \cite{Vilenkin:1983xq,Linde:1986fd,Linde:1995ck,
Vilenkin:1995yd,Guth:2000ka}.
That is, once inflation begins, even if some regions cease to
inflate there will always be an inflating region with increasing
physical volume.  This property holds in most models of inflation
that we can construct.  It relies on the fact that the scalar
inflaton field doesn't merely follow its classical equations
of motion, but undergoes quantum fluctuations, which can make
it temporarily roll up the potential instead of down.  The
regions in which this happens will have a larger potential
energy, and therefore a larger expansion rate, and therefore
will grow in volume in comparison to the other regions.  One
can argue that this process guarantees that inflation never
stops once it begins.

We can therefore imagine that the universe approaches a 
steady state (at least statistically), in which it is 
described by a certain fractal dimension \cite{Linde:1994xx}.
(Unfortunately, it seems impossible to extend such a
description into the past, to achieve a truly steady-state
cosmology \cite{Borde:1996pt}.)  This means that the universe
on ultra-large scales, much larger than the current Hubble
radius, may be dramatically inhomogeneous and isotropic,
and even raises the possibility that different post-inflationary
regions may have fallen into different vacuum states and
experience very different physics than we see around us.
Certainly, this picture represents a dramatic alteration
of the conventional view of a single Robertson-Walker 
cosmology describing the entire universe.

Of course, it should be kept in mind that the arguments in
favor of eternal inflation rely on features of the
interaction between quantum fluctuations and the gravitational
field which are slightly outside the realm of things we claim
to fully understand.  It would certainly be interesting to 
study eternal inflation within the context of string theory.

\section{Stringy cosmology}
\label{sec:stringy}

There is too much we don't understand
about both cosmology and string theory to make statements about 
the very early universe in string theory with any confidence.
Even in the absence of confidence, however, it is still worthwhile
to speculate about different possibilities, and work towards
incorporating these speculations into a more complete picture.

\subsection{The beginning of time}

Not knowing the correct place to start, a simple guess 
might be the (bosonic, NS-NS part of the)
low-energy effective action in $D$ dimensions,
\be
  S = -{1\over{16\pi G_D}}\int d^Dx\, \sqrt{-g} e^{-\phi}
  \left(R + \partial_\mu\phi \partial^\mu\phi - {1\over {12}}
  H_{\mu\nu\rho} H^{\mu\nu\rho} \right)\ ,
  \label{leea}
\ee
where $R$ is the Ricci scalar, $\phi$ is the dilaton, and
$H_{\mu\nu\rho} = \partial_{[\mu}B_{\nu\rho]}$ is the field
strength tensor for the two-form gauge field $B_{\nu\rho}$ (which
is typically set to zero in papers about cosmology).  
The existence of the dilaton implies
that the theory of gravity described by this action is a
scalar-tensor model (reminiscent of Brans-Dicke theory), not
pure general relativity.  Of course there are good experimental
limits on scalar components to the gravitational interaction, but
they are only sensitive to low-mass scalars ({\it i.e.}, 
long-range forces), so that the dilaton could escape detection
if we added a potential $V(\phi)$ to (\ref{leea}) which led
to a mass $m_\phi > 10^{-4}$~eV$\sim (10^{-1}$~cm$)^{-1}$.
As we've discussed, it is natural to hope that supersymmetry 
breaking induces a mass $m_\phi \sim 10^3$~GeV, so we would seem
quite safe.

So we really should include a potential for $\phi$ (not to mention
one for $B_{\nu\rho}$) in (\ref{leea}), but let's neglect it
for now and move boldly forward.  With this action, cosmological
solutions of the form
\be
  ds^2 = -dt^2 + \sum_{i=1}^{D-1} a_i^2(t) dx_i^2
\ee
(homogeneous but not necessarily isotropic) have a ``scale-factor
duality'' symmetry; for any solution $\{ a_i(t), \phi(t) \}$,
there is also a solution with
\be
  a_i' = {1\over{a_i}}\ ,\quad \phi' = \phi
  - 2\sum_i \ln a_i\ .
\ee
Thus, expanding solutions are dual to contracting solutions.
(In fact this is just $T$-duality, and 
is a subgroup of a larger $\o{D-1,D-1}$
symmetry.)  What is more, solutions with decreasing curvature
are mapped to those with increasing curvature.

This feature of the low-energy string action has led to
the development of the ``Pre-Big-Bang Scenario'', in which
the universe starts out as flat empty space, begins to
{\sl contract} (with increasing curvature), until reaching
a ``stringy'' state of maximum curvature, and then expands
(as curvature decreases) and commences standard cosmological
evolution \cite{Gasperini:1993em,prebb,lwc}.  (For related
considerations outside the Pre-BB picture, see \cite{Barrow:1998wd}.)

There are various questions about the Pre-BB scenario.  One is
a claim that significant fine-tuning is required in the initial
phase, in the sense that any small amount of curvature will 
grow fantastically during the evolutionary process and must
be extremely suppressed \cite{Turner:1997ih,Kaloper:1998eg}.  
Another is the role of the potential
for the dilaton.  We cannot set this potential to zero
on the grounds that the relevant temperatures are much higher than
the \SUSY-breaking scale $T_{\rm SUSY}\sim 10^3$~GeV; supersymmetry
is an example of a symmetry which is {\it not} restored at 
high temperatures \cite{Dine:1995uk}.  
Indeed, almost any state breaks supersymmetry.
In a thermal background, this breaking is manifested most
clearly by the differing occupation numbers for bosons and
fermions.  More generally, the \SUSY algebra
\be
  \{ Q, \bar{Q} \} = H + Z\ ,
\ee
with $Z$ a central charge,
implies that $Q\neq 0$ whenever $H\neq 0$, except in BPS
states, which feature a precise cancellation between $H$
and $Z$.  In the real world (in contrast to the world of
{\tt hep-th}) these are a negligible fraction of all
possible states.  It is not clear how \SUSY breaking
affects the Pre-BB idea.

Perhaps more profoundly, it seems perfectly likely that
the appropriate description of the high-curvature stringy
phase will be nothing like a smooth classical spacetime.
Evidence for this comes from matrix theory, not to mention
attempts to canonically quantize general relativity. 

There are other, non-stringy, approaches to the very
beginning of the universe, and it would be interesting to
know what light can be shed on them by string theory.
One is ``quantum cosmology'', which by some definitions
is just the study of the wave function of the universe,
although in practice it has the connotation of minisuperspace
techniques (drastically truncating the gravitational degrees
of freedom and quantizing what is left)
\cite{Hartle:1983ai,Linde:1995ck,Vilenkin:1998rp,Banks:1999ay}.
There is also the related idea of creation of baby universes
from our own \cite{Farhi:1990yr,Fischler:1990pk}.  
This is in principle a conceivable scheme,
as closed universes have zero total
energy in general relativity.  There is also the hope that
string theory will offer some unique resolution to the
question of cosmological (and other) singularities; studies
to date have had some interesting results, but we don't
know enough to understand the Big Bang singularity of
the real world \cite{larsenwilczek,maggioreriotto,Banks:1998vs}.

\subsection{Extra dimensions and compactification}

Of all the features of string theory, the one with the most
obvious relevance to cosmology is the existence of (6? 7?)
extra spatial (temporal?) dimensions.  The success of our
traditional description of the world as a (3+1)-dimensional
spacetime implies that the extra dimensions must be somehow
inaccessible, and the simplest method for hiding them is
compactification --- the idea that the extra dimensions
describe a compact space of sufficiently small size that they
can only be probed by very high energies.

Of course in general relativity (and even in string theory)
spacetime is dynamical, and it would be natural to expect the
compact dimensions to evolve.  However, the parameters
describing the size and shape of the compact dimensions
show up in our low-energy world as moduli fields whose values
affect the Standard Model parameters.  As discussed 
earlier, we have good limits on
any variation of these parameters in spacetime, and typically
appeal to \SUSY breaking to fix their expectation values.  This
raises all sorts of questions.  Why are three dimensions
allowed to be large and expanding while the others are small
and essentially frozen?  What is the precise origin of the 
moduli potentials?  What was the behavior of the extra 
dimensions in the early universe?

For the most part these are baffling questions, although there
have been some provocative suggestions.  One is by Brandenberger
and Vafa, who attempted to understand the existence of three
macroscopic spatial dimensions in terms of string dynamics
\cite{Brandenberger:1989aj}.
Consider an $n$-torus populated by both momentum modes
and winding modes of strings.
The momentum and winding modes are dual to each other under
$T$-duality ($R \rightarrow 1/R$), and have opposite effects
on the dynamics of the torus: the momentum modes tend to make 
it expand, and the winding modes to make it contract.  (It's
counterintuitive, but true.)  We can therefore have a static
universe at the self-dual radius where the two effects are
balanced.  However, when wound strings intersect they tend to
intercommute and therefore unwind.
Through this process, the balance holding the torus at the
self-dual radius can be upset, and the universe will begin
to expand, hopefully evolving into a conventional Friedmann
cosmology.

But notice that in a sufficiently large number of spatial
dimensions, one-dimensional strings will generically never
intersect.  (Just as zero-dimensional points will generically
intersect in one dimension but not in two or more dimensions.)
The largest number in which they tend to intersect is three.
So we can imagine a universe that begins as a tiny torus
in thermal equilibrium at the self-dual point, until some
winding modes happen to annihilate in some three-dimensional
subspace which then begins to expand, forming our universe.
Of course a scenario such as this loses some of its charm
in a theory which has not only strings but also 
higher-dimensional branes.  (Not to mention that toroidal 
compactifications are not pheonomenologically favored.)

An alternate route is to take advantage of the existence of
these branes, by imagining that we are living on one.  That is to
say, that the reason why the extra dimensions are invisible to
us is not simply because they are so very small that low-energy
excitations cannot probe them, but because we are confined to
a three-dimensional brane embedded in a higher-dimensional
space.  We know that we can easily construct field theories
confined to branes, for example a $\u{N}$ gauge theory by
stacking $N$ coincident branes; it is not an incredible
stretch to imagine that the entire Standard Model can be
constructed in such a way (although it hasn't been done yet).
Unfortunately, it seems impossible to entirely do away with
the necessity of compactification, since there is one force
which we don't know how to confine to a brane, namely gravity
(although see below).

We therefore imagine a world in which the Standard Model
particles are confined to a three-brane, with gravity
propagating in a higher-dimensional ``bulk'' which includes
compactified extra dimensions.  In $D$ spacetime dimensions,
Newton's law of gravity can be written
\be
  F_{(D)}(r) = \widetilde{G}_{(D)} {{m_1 m_2}\over{ r^{D-2}}}\ ,
  \label{gauss}
\ee
where $\widetilde{G}_{(D)}$ is the $D$-dimensional Newton's
constant with appropriate factors of $4\pi$ absorbed.
If we compactify $D-4$ of the spatial dimensions on a compact
manifold of volume $V_{(D-4)}$, the effective 4-dimensional
Newton's constant is
\be
   \widetilde{G}_{(4)}\sim {{\widetilde{G}_{(D)}}\over{V_{(D-4)}}}\ .
\ee
We can rewrite this in terms of what we will define as the
Planck scale, $M_\P = \widetilde{G}_{(4)}^{-1/2}$, and the
"fundamental'' scale, $M_* = \widetilde{G}_{(D)}^{-1/(D-2)}$,
as
\be
  M_\P^2 \sim M_*^{D-2} V_{(D-4)}\ .
\ee
In conventional compactification, $M_*\sim M_\P$ and
$V_{(D-4)}\sim M_\P^{-(D-4)}$, so this relation is straightforwardly
satisfied.  But we can also satisfy it by lowering the fundamental
scale and increasing the compactification volume.  Imagine
that the compactification manifold has $n$ ``large'' dimensions
of radius $R$ and $D-4-n$ dimensions of radius $M_*^{-1}$.
Then 
\be
  R \sim \left({{M_\P}\over{M_*}}\right)^{2/n} M_*^{-1}\ .
\ee
A scenario of this type was proposed by Horava and Witten \cite{mtheory},
who suggested that the gravitational coupling could unify with the
gauge couplings of GUT's by introducing a single large extra dimension
with $R\sim (10^{15}~{\rm GeV})^{-1}$.  

But we can go further.  The lowest value we can safely imagine the 
fundamental
scale having is $M_*\sim 10^3$~GeV; otherwise we would have
detected quantum gravity at Fermilab or CERN.  This value is
essentially the desired low-energy supersymmetry breaking scale
({\it i.e.} just above the electroweak scale), so it is tempting to
explain the apparent hierarchy $M_*/M_{\rm EW}\sim 10^{15}$
by trying to move $M_*$ all the way down to $10^3$~GeV
\cite{Antoniadis:1990ew,Lykken:1996fj,Arkani-Hamed:1998rs,
Antoniadis:1998ig}.
(Note that supersymmetry itself can stabilize the
hierarchy, but doesn't actually explain it.)  Then we have
\be
  R\sim 10^{30/n - 3}~{\rm GeV}^{-1} \sim
  10^{30/n - 17}~{\rm cm}\ .
\ee
For $n=1$, we have a single extra dimension of radius
$R\sim 10^{13}$~cm, about the distance from the Sun to the
Earth.  This is clearly ruled out, as such a scenario predicts
that gravitational forces would fall off as $r^{-3}$ for distances
smaller than $10^{13}$~cm.  But for $n=2$ we have $R\sim
10^{-2}$~cm, which is just below the limits on deviations from
the inverse square law from laboratory experiments.  Larger
$n$ gives smaller values of $R$; these are not as exciting
from the point of view of having macroscopically big extra
dimensions, but may actually be the most sensible from a
physics standpoint.

So we have a picture of the world as a 3-brane with Standard
Model particles restricted to it, and gravity able to propagate
into a bulk with extra dimensions which are compactified but
perhaps of macroscopic size, with a fundamental scale 
$M_*\sim 10^3$~GeV and the observed Planck scale simply
an artifact of the large extra dimensions.  (There is still
something of a hierarchy problem, since $R$ must be larger 
than $M_*$ to get the Planck scale right.)  Such
scenarios are subject to all sorts of limits from astrophysics
and accelerator experiments, from processes such as gravitons
escaping into the bulk.  (In these models gravity becomes
strongly coupled near $10^3$~GeV.)

There are also going to be cosmological implications,
although it is not precisely clear as yet what these
are (see for example \cite{Arkani-Hamed:1998nn,Arkani-Hamed:1999kq,
Lykken:1998fb,Banks:1999eg,Hall:1999mk,Binetruy:1999ut},
not to mention many papers subsequent to the writing of
these notes).  
Our entire discussion of the thermal history of the 
universe for $T > 10^3$~GeV would obviously need to be
discarded.  Baryogenesis will presumably be modified.  Inflation
is a very interesting question, including the issues of
inflation in the bulk vs. inflation in the boundary.  Of 
course we don't know what stabilizes the large extra dimensions,
but then again we don't know much about moduli stabilization
in conventional scenarios.  There is also the interesting
possibility of a 3-brane parallel to the one we live on,
which only interacts with us gravitationally, and on which
the dark matter resides.  There are some cosmological problems,
though --- most clearly, the issue of why the bulk is not highly
populated by light particles that one might have expected to
be left over from an early high-temperature state; presumably
reheating after inflation cannot be to a very high temperature
in these models (although we must at the very least have
$T_{\rm reheat} > 1$~MeV to preserve standard nucleosynthesis).
Clearly there is a good deal of work left to do in exploring
these scenarios.

After these notes were written, Randall and Sundrum 
\cite{randallsundrum} found a loophole in the conventional
wisdom that gravity cannot be confined to a brane.  They showed
that a single extra dimension could be infinitely large, but
still yield an effective 4-dimensional gravity theory on the
brane, if the bulk geometry were anti-de~Sitter rather than
flat.  The curvature in the extra dimension can then effectively
confine gravity to the vicinity of the brane.  In the subsequent
months a great deal of effort has gone into understanding
cosmological and other ramifications of Randall-Sundrum 
scenarios, which are surely worthy of their own review article
by this point.

\subsection{The late universe}

The behavior of gravity and particle physics on extremely
short length scales and high energies is largely uncharted
territory, and it is clear that string theory, if correct,
will play an important role in understanding this regime.
But it is also interesting to contemplate the possibility
of new physics at ultra-large length scales and low energies.
You might guess that experiments in the zero-energy limit
are straightforward to perform, but in fact it requires great
effort to isolate yourself from unwanted noise sources in
this regime.  Cosmology offers a way to probe physics on
the largest observable length scales in the universe, and it
is natural to take advantage.

We spoke in Section \ref{subsec:scalefactor} about the apparent 
acceleration of the 
universe, which, if verified, would be a dramatic indication
of new physics at very low energies.  Explaining the 
observations with a positive vacuum energy $\rho_{\rm V}
= M_{\rm V}^4$ requires
\be
  M_{\rm V}\sim 10^{-3}~{\rm eV}\ ,
\ee
which is remarkably small in comparison to $M_{\rm SUSY}\sim
10^3$~GeV\ $= 10^{12}$~eV, not to mention $M_\P \sim
10^{18}$~GeV\ $= 10^{27}$~eV.  It does, of course, induce
the irresistible temptation to write
\be
  M_{\rm V}\sim {{M_{\rm SUSY}^2}\over{M_\P}}\ .
\ee
This is a numerological curiosity without a theory that
actually predicts it, although it has the look and feel of
similar relations familiar from models in which \SUSY breaking
is communicated from one sector to another by gravitational
interactions.  Another provocative relation is
\be
   M_{\rm V}\sim e^{-1/2\alpha}M_\P\ ,
\ee
where $\alpha$ is the fine-structure constant.  Again, it
falls somewhat short of the standards of a scientific theory,
but it does suggest the possibility that the vacuum energy
would be precisely zero if it were not for some small 
nonperturbative effect.  There may even be ways to get such
effects in string theory \cite{harvey}.

More generally, we can classify vacuum energy as coming
from one of three
categories:  ``true vacua'', which are global minima of the
energy density; ``false vacua'', which are local but not
global minima; and ``non-vacua'', which is a way of expressing
the idea that we have not yet reached a local minimum value
of the potential energy.  For example, we could posit the
existence of a scalar field $\phi$ with a very shallow potential
\cite{rp,fhsw,tw,cds,Steinhardt:1999nw,carroll00}.  
From our analysis in Section \ref{subsec:relicabs}, the field
will be overdamped when $V'' < H$, and its potential energy
will dominate over its kinetic energy (exactly as in inflation).
Such a possibility has been dubbed ``quintessence''.
For a quintessence field to explain the accelerating universe,
it must have an effective mass
\be
  m_\phi \equiv \sqrt{V''} \leq H_0 \sim 10^{-33}~{\rm eV}\ ,
\ee
and a typical range of variation over cosmological timescales
\be
  \Delta\phi \sim M_\P \sim 10^{18}~{\rm GeV}\ .
\ee
From a particle-physics point of view, these parameters seem
somewhat contrived, to say the least.  In fact, the same
fifth-force experiments and variation-of-constants limits
that we previously invoked to argue against the existence of massless
moduli are applicable here, and point toward the necessity
of some additional structure in a quintessence theory in
order to evade these bounds \cite{carroll}.  However, 
quintessence models have the benefit of involving dynamical
fields rather than a single constant, and it may be possible to
take advantage of these dynamics to ameliorate the ``coincidence
problem'' that $\Omega_\Lambda\sim \Omega_{\rm M}$ today
(despite the radically different time dependences of these
two quantities).  In addition, there
may be more complicated ways to get a time-dependent vacuum
energy that are also worth exploring \cite{ac,Bucher:1998mh}.
The moduli fields of string theory could provide potential
candidates for quintessence, and the acceleration of the
universe more generally provides a rare opportunity for string
theory to provide an explanation of an empirical fact.

We could also imagine that string theory may have more profound
late-time cosmological consequences than simply
providing a small vacuum energy or ultralight scalar fields.
An interesting move in this direction is to explore the 
implications of the ``holographic principle'' for cosmology.
This principle was inspired by our semiclassical expectation
that the entropy of a black hole, which in traditional statistical
mechanics is a measure of the number of degrees of freedom in
the system, scales as the area of the event horizon rather
than as the enclosed volume (as we would expect the degrees
of freedom to do in a local quantum field theory).  In its
vaguest (and therefore most likely to be correct) form, the
holographic principle proposes that a theory with gravity in
$n$ dimensions (or a state in such a theory) is equivalent in 
some sense to a theory without gravity in $n-1$ dimensions
(or a state in such a theory).  Making this statement more
precise is an area of active investigation and controversy;
see Susskind's lectures for a more complete account
\cite{bigsuss}.
The only context in which the holographic equivalence has been
made at all explicit is in the AdS/CFT correspondence, where
the non-gravitational theory can be thought of as living on
the spacelike boundary at conformal infinity of the AdS
space on which the gravitational theory lives.  

Regrettably, we don't live in anti-de~Sitter space, which
corresponds to a RW metric with a negative cosmological
constant and no matter, since our universe seems to feature
both matter and a positive cosmological constant.  How
might the holographic principle apply to more general
spacetimes, without the properties of conformal infinity
unique to AdS, or for that matter without any special
symmetries?  A possible answer has been suggested by
Bousso \cite{bousso}, building upon ideas of Fischler and
Susskind \cite{Fischler:1998st}.  The basic idea is to
the area $A$ of the boundary of a spatial volume to the amount
of entropy $S$ passing through a certain null sheet bounded by that
surface.  (For details of how to construct an appropriate
sheet, see the original references.)  Specifically, the
conjecture is that
\be
  S \leq A/4G\ .
\ee
This is more properly an entropy bound, not a claim about
holography; however, it seems to be a short step from limiting
entropy (and thus the number of degrees of freedom) to 
claiming the existence of an underlying theory dealing directly
with those degrees of freedom.

Does this proposal have any consequences for cosmology?
It is straightforward to check that the bound is satisfied
by standard cosmological 
solutions\footnote{It is amusing to note that
\cite{Fischler:1998st} underestimated
the entropy in the current universe, which Penrose \cite{penrose}
has pointed out is dominated not by photons but by massive black
holes at the centers of galaxies, which have total entropy
$S\sim 10^{100}$.  The estimate was therefore off by a
factor of $10^{14}$, although the conclusions are left
unchanged.  (This is a common occurrence in cosmology.)},
and a classical version can even be proven to hold under
certain assumptions \cite{Flanagan:2000jp}.  One optimistic
hope is that holography could be responsible for the small
observed value of the cosmological constant
(see for example \cite{Banks:1995uh,Cohen:1998zx,Horava:2000tb,
Banks:2000fe,Thomas:2000km}).  Roughly speaking, this hope is
based on the idea that there are far fewer degrees of freedom
per unit volume in a holographic theory than local quantum field
theory would lead us to expect, and perhaps the unwarranted
inclusion of these degrees of freedom has been leading to an
overestimate of the vacuum energy.  It remains to be seen 
whether a workable implementation of this idea can capture
the successes of conventional cosmology.

\section{Conclusions}
\label{sec:conclusions}

The last several years have been a very exciting time in
string theory, as we have learned a great deal about
non-perturbative aspects of the theory, most impressively the
dualities connecting what were thought to be different theories.
They have been equally exciting in cosmology, as a wealth of
new data have greatly increased our knowledge about the 
constituents and evolution of the universe.  The two subjects
still have a long way to go, however, before their respective
domains of established understanding are definitively 
overlapping.  One road toward that goal is to work diligently
at those aspects of string theory and cosmology which are best
understood, hoping to enlarge these regions until they someday
meet.  Another strategy is to leap fearlessly into the murky
regions in between, hoping that our current fumbling
attempts will mature into more solid ideas.  Both approaches are,
of course, useful and indeed necessary; hopefully these notes 
will help to empower the next generation of fearless leapers.

\section{Acknowledgments}

I've benefited from conversations with many colleagues,
including Tom Banks, Alan Guth, Gary Horowitz, 
Steuard Jensen, Clifford Johnson, Finn Larsen,
Donald Marolf, Ue-Li Pen, Joe Polchinski, Andrew Sornborger,
Paul Steinhardt, and Mark Trodden, as well as numerous participants 
at TASI-99.  I would like to thank the organizers (Jeff Harvey,
Shamit Kachru and Eva Silverstein) for
arranging a very stimulating school, and the participants for
their enthusiasm and insight.
This work was supported in part by the National Science Foundation
under grant PHY/94-07195, the U.S. Department of Energy, and the
Alfred P. Sloan Foundation.

\end{document}